\documentclass[a4paper, aps, prx, twocolumn, superscriptaddress]{revtex4-2}

\usepackage{braket}
\usepackage{graphicx}
\usepackage{amsmath}
\usepackage{amssymb}
\usepackage{hyperref}
\usepackage[all]{hypcap}
\hypersetup{colorlinks=True,allcolors=blue}
\usepackage{xcolor}
\usepackage{comment, nicefrac}
\usepackage{bm}
\usepackage{mathtools}  
\usepackage{natbib}
\usepackage{siunitx}
\DeclareSIUnit{\arbitraryunits}{arb.u.}
\DeclareSIUnit{\belmilliwatt}{Bm}
\newcommand*{\e}{\ensuremath{\mathrm{e}}}
\newcommand{\imu}{\ensuremath{\mathrm{i}}}

\newcommand{\WMIaff}{\affiliation{Walther-Meissner-Institut, Bayerische Akademie der Wissenschaften, 85748 Garching, Germany}}
\newcommand{\TUMaff}{\affiliation{Technical University of Munich, TUM School of Natural Sciences, Physics Department, 85748 Garching, Germany}}
\newcommand{\UniBwaff}{\affiliation{University of the Bundeswehr Munich, Professorship of Electromagnetic Theory and Photonics, 85579 Neubiberg, Germany}}
\newcommand{\MCQSTaff}{\affiliation{Munich Center for Quantum Science and Technology (MCQST), 80799 Munich, Germany}}
\newcommand{\PQaff}{\affiliation{Peak Quantum GmbH, 85748 Garching, Germany}}
\newcommand{\MQVaff}{\affiliation{Munich Quantum Valley gGmbH, 85748 Garching, Germany}}

\begin{document}

\title{Field-induced phase-matching in Josephson traveling wave parametric amplifiers}

\author{Daniil~E.~Bazulin}\WMIaff\TUMaff
\author{Simon~Gandorfer}\WMIaff\TUMaff 
\author{Niklas~Bruckmoser}\WMIaff\TUMaff
\author{Kedar~E.~Honasoge}\TUMaff\PQaff 
\author{Leon~Koch}\TUMaff\PQaff 
\author{Maria-Teresa~Handschuh}\WMIaff\TUMaff 
\author{Matthias~Hirschbichler}\UniBwaff
\author{Kirill~Shulga}\affiliation{International Center for Elementary Particle Physics,~The University of Tokyo, 7-3-1 Hongo, Bunkyo-ku, Tokyo 113-0033, Japan} 
\author{Achim~Marx}\WMIaff 
\author{Michael~Haider}\UniBwaff 
\author{Stefan~Filipp}\WMIaff\TUMaff\MCQSTaff\MQVaff  
\author{Kirill~G.~Fedorov}\WMIaff\TUMaff\MCQSTaff 

\date{\today}

\begin{abstract}
    Superconducting Josephson traveling wave parametric amplifiers (JTWPAs) enable broadband, quantum-limited amplification of weak microwave signals. JTWPAs are essential for the single-shot readout of superconducting qubits, broadband vacuum squeezing, or the generation of microwave cluster states. Phase-matching between signal and pump modes in such JTWPAs represents a major engineering challenge, which is typically solved by integrating resonators in a nonlinear medium in order to tailor electromagnetic dispersion relations and achieve a broadband, high-gain amplification. Here, we experimentally demonstrate a phase-matching approach using a homogeneous nonlinear medium based on Josephson elements by exploiting a magnetic field inhomogeneity. We show how the latter may lead to the sign reversal of the 3-wave mixing nonlinearity and stabilization of the parametric amplification. We envision this technique as a viable alternative to conventional phase-matching approaches.
\end{abstract}

\maketitle

\section{Introduction}

Quantum-limited amplification of weak microwave signals is of paramount importance for a variety of fundamental and applied  objectives alike, such as the search of dark matter axions~\cite{Bartram2023} or qubit readout in scalable superconducting quantum processors~\cite{Bravyi2022,Martinis2025}. Modern Josephson traveling wave parametric amplifiers (JTWPAs) are capable of fulfilling many of the related stringent requirements including high-gain amplification, large instantaneous bandwidth, high compression power, and quantum-limited noise performance, among others. Further novel applications of JTWPAs may include the generation of broadband squeezed states~\cite{Esposito2022, Qiu2023} or photonic cluster states~\cite{Pfister2020,Lingua2025} for continuous-variable quantum information processing and microwave quantum communication~\cite{Fedorov2021,Fesquet2023}. 

\begin{figure*}[t]
    \centering
    \includegraphics[width=1.0\linewidth]{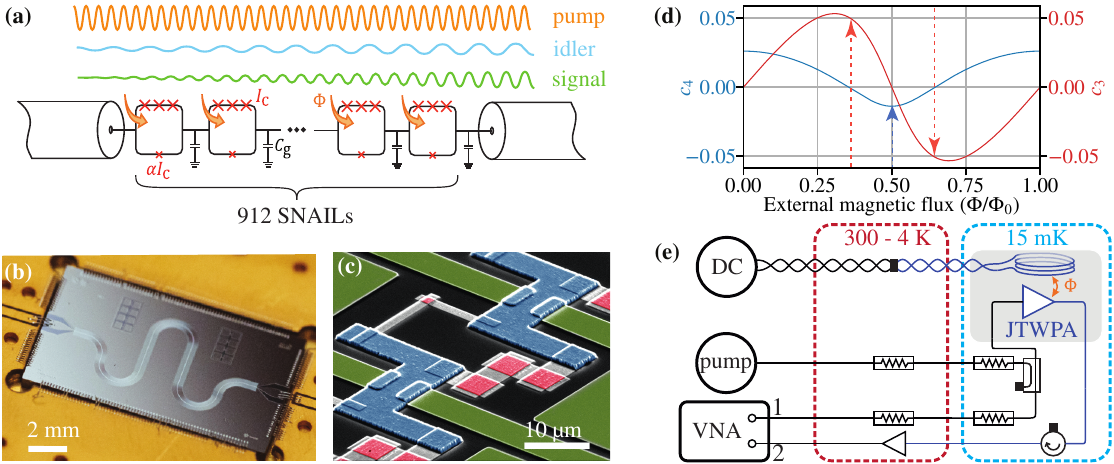}
    \caption{JTWPA description and measurement schematics. (a) Circuit diagram of the JTWPA consisting of a chain of 912 SNAILs with the critical current of large Josephson junctions $I_\mathrm{c}$ and the critical current of small Josephson junctions $\alpha I_\mathrm{c}$, where $\alpha\simeq\num{0.12}$. Capacitance to ground is $C_\mathrm{g}$. The SNAIL chain is biased with the external magnetic flux $\Phi$. (b) Photo image of the bonded JTWPA sample. We use fishbone capacitors to match the JTWPA impedance to $\qty{50}{\ohm}$. (c) False-colored SEM image of a single SNAIL circuit. Green color corresponds to niobium making up the ground plane and fishbone capacitors. Red color marks Al/AlO\textsubscript{x}/Al Josephson junctions in the double-mask Manhattan-style overlap geometry. Blue color marks aluminum bandages providing a galvanic connection between different superconducting layers. (d) Dependence of 3-wave $c_3$ and 4-wave $c_4$ nonlinear coefficients of a single SNAIL with $\alpha=\num{0.12}$ as a function of the external magnetic flux. Red (blue) dashed line corresponds to a sweet spot for the 3WM (4WM) processes with $c_4 = 0$ ($c_3 = 0$). (e) Simplified schematic of the main experimental setup. Here, the JTWPA sample holder is located at the millikelvin stage of a dilution  fridge with an attached superconducting coil within an aluminum magnetic shielding (gray outline). We encode superconducting cables with blue color and normal cables with black color. For microwave transmission characterization measurements we use a vector network analyzer (VNA) and a microwave vector signal generator as the pump source.}
    \label{fig:intro}
\end{figure*}

JTWPAs consist of a chain of elements based on Josephson junctions which allow for a wave mixing between co-propagating signal and pump microwave modes, ultimately enabling the parametric amplification process. One of the most popular choices for such circuits is the superconducting nonlinear asymmetric inductive element (SNAIL), as shown in Fig.\,\ref{fig:intro}, which allows one to use both the 3-wave mixing (3WM) and 4-wave mixing (4WM) processes depending on the chosen flux bias point and SNAIL parameters~\cite{Frattini2017}. 

Phase-matching of propagating modes participating in the nonlinear mixing process is essential for the efficiency of the amplification process. However, a phase mismatch may occur due to a linear dispersion in transmission lines as well as due to a self- and cross-phase modulation in nonlinear media. There are numerous techniques that help to establish an optimal balance between phase velocities of the signal and pump modes co-propagating through the JTWPA nonlinear medium. In particular, the resonant phase-matching approach employs the integration of lumped element microwave resonators in the JTWPA nonlinear medium in order to modify the dispersion relation for the propagating modes and avoid destructive interference due to accumulated phase mismatches~\cite{O’Brien2014, Macklin2015, White2015, Perelshtein2022}. Alternative phase-matching techniques involve the modulation of various JTWPA parameters~\cite{Planat2020, Chang2025, FadaviRoudsari2023}, exploiting a sign reversal of the Kerr nonlinearity~\cite{Ranadive2022, Bell2026}, or using opposing directions of phase and group velocities~\cite{Kow2025} to achieve the same goal. 
 
Depending on the type of parametric pumping, one can realize 3WM or 4WM processes in the JTWPA. Historically, the 4WM process was the first to be experimentally demonstrated, since Josephson junctions without any external bias possess a symmetric energy profile whose leading nonlinearity is cubic in the current--phase relation\,\cite{Macklin2015}. Such a medium allows efficient mixing of four microwave photons, which corresponds to a down-conversion process of two pump photons into a pair of signal and idler photons, placing the pump frequency into the center of the amplification bandwidth. In contrast, the pure 3WM process requires pumping at twice the desired center frequency. This approach removes undesirable spectral features associated with the strong pump tone and its phase-matching structures around the center amplification frequency. Pure 3WM does not suffer from  the self- and cross-phase modulation, which implies an easier phase-matching, a more efficient energy transfer, and higher compression powers of the corresponding JTWPAs~\cite{Zorin2016}. However, these arguments also hold for other, parasitic, up-conversion processes, such as the second harmonic and sum frequency generation. Enabling the 3WM process requires symmetry breaking of the potential energy landscape, which necessitates an external field bias. Experimentally, it can be implemented either by applying a direct current bias~\cite{Gaydamachenko2025, Livreri2024} or by using a magnetic field bias~\cite{Planat2020, Ranadive2022, FadaviRoudsari2023, DiVora2023, Kow2026} applied to the JTWPA medium. 

In this manuscript, we focus on the experimental demonstration of phase-matching in a SNAIL-based JTWPA by exploiting an inhomogeneous distribution of the magnetic bias field caused by the field focusing effect in superconducting circuits as well as by the inherent geometric field broadening due to the finite dimensions of the coil used for external flux biasing. We show that we can achieve phase-matching in the 3WM process using a simple design of a uniform nonlinear transmission line without any on-chip phase-matching techniques.

\section{Device and methods}
\label{sec:device}

We use SNAILs as a basic building block, because their nonlinearity order can be tuned \textit{in situ} by applying an external magnetic field~\cite{Zapata1996, Frattini2017}. Corresponding nonlinear coefficients can be derived from the Taylor expansion of the SNAIL potential energy\,\cite{Frattini2017, Ranadive2022, Haider2024}
\begin{equation}
    \begin{split}
       	c_3 = \frac{1}{2} \left[ \alpha \sin{\varphi^*} + \frac{1}{N^2} \sin\left(\frac{\varphi^* - \varphi}{N}\right) \right]\,, \\
        c_4  = \frac{1}{6} \left[ \alpha \cos{\varphi^*} + \frac{1}{N^3} \cos\left(\frac{\varphi^* - \varphi}{N}\right) \right]\,,
    \end{split}
    \label{eq:SNAILs_non_lin_coeffs}
\end{equation}
where $\alpha < 1$ is the asymmetry parameter, $N$ is the number of large junctions, $\varphi^*$ is the metastable Josephson phase across an individual SNAIL, and $\varphi = 2 \pi\Phi / \Phi_0$ is a phase bias expressed via the external magnetic flux $\Phi$ normalized to the flux quantum $\Phi_0$. Here, we neglect the geometric inductance of the SNAIL loop. Depending on $\alpha$ and $N$, the SNAIL can be tuned to sweet spots, where it exhibits a pure quadratic ($c_3 \neq 0, c_4 = 0$) or pure Kerr nonlinearity ($c_3 = 0, c_4 \neq 0$), as shown in Fig.\,\ref{fig:intro}(d)~\cite{Sivak2019}. 

\begin{figure*}
    \centering
    \includegraphics[width=1.0\linewidth]{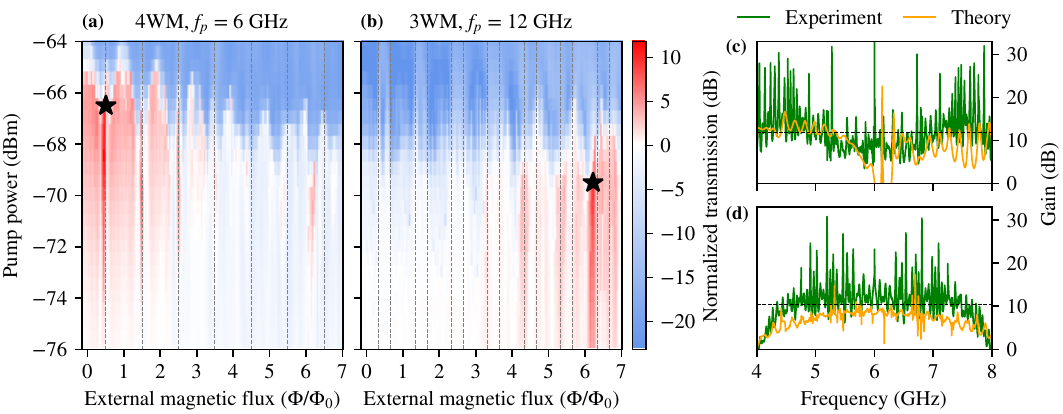}
    \caption{Average parametric gain in the range \qtyrange[range-phrase=~--~]{4}{8}{\giga\hertz} as a function of the external applied magnetic flux $\Phi$ through the SNAILs and the pump power at the JTWPA input. Panel (a) illustrates the 4WM regime with the pump frequency set at \qty{6.0}{\giga\hertz}. Dashed lines mark theoretical 4WM sweet spots at $(\num{0.5} + m) \Phi_0$, where $m$ is an integer. Panel (b) illustrates the 3WM regime with the pump frequency set at \qty{12.0}{\giga\hertz}. Dashed lines mark theoretical 3WM sweet spots $(\pm\num{0.354} + m) \Phi_0$, where the SNAILs have a pure quadratic nonlinearity, $c_3 \neq 0$ and $c_4=0$. Panels (c) and (d) show exemplary gain profiles (in green), corresponding to working points marked with stars in the respective  plots (a) and (b), together with theoretical expectation profile (in orange) based on multi-mode model (see Section \ref{sec:model}). Dashed lines indicate the mean gain value averaged over the range \qtyrange[range-phrase=~--~]{4}{8}{\giga\hertz}.}
    \label{fig:3WM_4WM_avg_gain_maps}
\end{figure*}

We fabricate our JTWPA using a four-mask process, similar to the process developed for the fabrication of superconducting processors~\cite{Roy2025} and narrow-band, quantum-limited Josephson parametric amplifiers~\cite{Honasoge2025}. We use a coplanar waveguide geometry to create a nonlinear transmission line consisting of $N_\mathrm{cells} = \num{912}$ SNAILs. The ground plane, microwave contact pads, and fishbone capacitors are made out of niobium~\cite{Lu2022, Chaudhuri2017, Malnou2021, Chang2025}. The fishbone capacitors are placed at each SNAIL cell and are important for matching the JTWPA characteristic impedance to $50\,\Omega$ for a selected working point defined by the external magnetic field. As the basic cell, we use a SNAIL with the number of big junctions $N = 3$ and asymmetry coefficient $\alpha=\num{0.12}$. These SNAILs employ the double-mask Manhattan-style overlap Josephson junctions fabricated in a two-step electron-beam lithography~\cite{Wang2025, Mahboob2026}. This approach allows to avoid many limitations related to the conventional shadow evaporation process,  maintains a very high yield of Josephson junctions (nearly $100\%$), and allows to rotate the SNAIL circuits by an arbitrary angle in order to implement meandered coplanar waveguide (CPW) structures, as shown in Fig.\,\ref{fig:intro}(b). In this process, we first evaporate the bottom electrode of a Josephson junction, followed by a lift-off process, spin-coating, and patterning of the top electrode. Afterwards, we use argon milling in order to remove native oxides formed during the exposure to ambient air and to form a tunnel oxide barrier for the Josephson junction in a controlled manner by purging oxygen. Afterwards, we evaporate the top aluminum electrode without breaking vacuum. Finally, we use an Al bandaging process~\cite{Dunsworth2017} to galvanically connect Josephson junctions to other superconducting leads, thus, forming the SNAIL loops and the entire JTWPA structure. Figure\,\ref{fig:intro} shows an SEM picture of an individual fabricated SNAIL (c) as well as the entire JTWPA wire-bonded to a printed circuit board for characterization measurements (b). This bonded chip is placed in a copper sample holder with a macroscopic superconducting coil mounted on top. The latter is used to control the external magnetic field through the JTWPA. In order to protect the sample from external magnetic noise, we put the entire assembly in an aluminum superconducting shield and cool it down to a temperature of \qty{15}{\milli\kelvin} using a commercial dilution refrigerator. Finally, we perform microwave transmission measurements using a room temperature vector network analyzer (VNA) and apply the pump tone via an additional directional coupler mounted in front of the sample assembly at the dilution stage level, as shown in Fig.\,\ref{fig:intro}(e).

\section{Experimental results}

We characterize the JTWPA performance by measuring scattering parameters $S_{21}$ in the frequency range of \qtyrange{4}{8}{\giga\hertz} with and without the pump tone as a function of the external magnetic flux. In the 4WM regime, we apply the continuous pump tone at \qty{6}{\giga\hertz}, while for the 3WM regime, we set the pump frequency to \qty{12}{\giga\hertz}. In order to simplify the experimental data analysis, we take an average over a normalized broadband response in terms of $S_{21}$ and plot it as a color map as a function of the external flux and pump power (see Fig.\,\ref{fig:3WM_4WM_avg_gain_maps}). Here, we observe moderate values of the broadband JTWPA gain in the 4WM regime around zero flux values and, as expected, higher gain values at optimal flux points corresponding to $(\num{0.5} + m) \Phi_0$, where $m$ is an integer, marked with dashed lines in Fig.\,\ref{fig:3WM_4WM_avg_gain_maps}(a). At these optimal working points the phase-matching condition is fulfilled due to the Kerr sign reversal~\cite{Ranadive2022}. Furthermore, we observe that the 4WM broadband gain regime is monotonically suppressed with an increasing flux offset and completely disappears for $m \gg 1$, as illustrated in Fig.\,\ref{fig:3WM_4WM_avg_gain_maps}(a).

In the 3WM regime, the amplification behavior of our JTWPA notably differs from 4WM, as shown in Fig.\,\ref{fig:3WM_4WM_avg_gain_maps}(b). Here, at low values of magnetic flux offset $\Phi \sim \Phi_0$, no amplification process is observed, even at the expected 3WM sweet spots marked by dashed lines. However, the situation changes when we shift the external flux offset to larger values, $\Phi \gtrsim 6 \Phi_0$, where the 3WM amplification process slowly emerges with increasing external magnetic flux $\Phi$ and reaches gain values similar to the 4WM regime.

One possible explanation for the absence of 3WM at low flux offsets is based on a second harmonic generation (SHG) of the pump mode, which represents a common loss channel in the 3WM regime~\cite{FadaviRoudsari2023, RenbergNilsson2023}. In order to test this hypothesis, we moved the device to a broadband cryogenic setup, where one can probe the transmission signal in a wider range of frequencies, specifically, \qtyrange{6}{20}{\giga\hertz}. Here, one would expect that around $\Phi \approx 0$ the pump mode at \qty{12}{\giga\hertz} generates the second harmonic, which strongly limits the 3WM amplification, while for higher values of magnetic flux $\Phi \gtrsim 6\Phi_0$ stronger inhomogeneous magnetic fields lower the cut-off frequency of the JTWPA due to the averaging of plasma frequencies of the SNAIL chain. Once the cut-off frequency is below the second pump harmonic, the SHG process becomes strongly suppressed, and the 3WM regime regains its amplification strength. However, the experimental broadband measurements show that the cut-off frequency is located around \qty{18}{\giga\hertz} at zero magnetic field, thus, disallowing the SHG of the pump mode from the very beginning. The second explanation is related to a similar mechanism: the 3WM amplification suppression may happen due to an unwanted photonic gap near the pump frequency of \qty{12}{\giga\hertz}, leading to the reflection of the pump mode. Once again, larger inhomogeneous magnetic fields may shift this bandgap to lower frequencies, allowing the pump mode to propagate through the JTWPA at larger flux offsets and enable the 3WM process. However, our broadband measurements also showed an absence of any flux-tunable resonance features in the vicinity of 3WM pump frequencies, ruling out this hypothesis as well.

Moreover, we investigate the 3WM regime at different pump frequencies and observe a similar pattern, as shown in Fig.\,\ref{fig:exp_theory_comparison}(a). Here, at low external magnetic fields, the 4WM process dominates, creating a broadband amplification response with the pump frequency at the center. As the external flux increases, one can see that the 4WM process fades out and the 3WM process appears. Notably, the latter emerges at proportionally lower external flux offsets as a function of the decreasing 3WM pump frequency, characterized by a corresponding threshold flux value, $\Phi_\mathrm{t}(f)$. These measurements identify the magnetic field distribution as an important parameter that can be used to control the device behavior. Intuitively, this inhomogeneity leads to a different phase bias of the SNAILs at different spatial locations and may substantially modify the JTWPA spectral properties. 

We also performed noise measurements of the device operated at the working points marked with stars in Fig.\,\ref{fig:3WM_4WM_avg_gain_maps}, resulting in $\num{0.32}\pm\num{0.27}$ and  $\num{0.45}\pm\num{0.23}$ photons on average above the standard quantum limit (SQL) for 4WM and 3WM operation regimes, respectively. Detailed explanation of the method and results can be found in App.\,\ref{app:noise}.

\section{Magnetic field inhomogeneity}
\label{sec:field_inhomogeneity}

As described in Section \ref{sec:device}, we use the external macroscopic coil made using a superconducting NbTi wire in order to control the magnetic field and corresponding flux bias through the SNAILs. This approach has a number of important experimental advantages, such as very low noise levels in the generated magnetic fields due to a large coil inductance and a broad range of applicable supercurrents. At the same time, due to the finite dimensions of the coil, a corresponding magnetic field distribution across the JTWPA cannot be considered perfectly homogeneous and its profile can be used as an additional degree of freedom. The distribution shape can be roughly estimated from finite element electromagnetic simulations, with additional contributions of the finite winding layer distribution and the mounting angle of the coil with respect to the sample holder.
Furthermore, the superconducting CPW geometry by itself represents another major source of the magnetic field inhomogeneity. Here, the Meißner effect in a superconducting ground plane surrounding the SNAIL chain leads to a noticeable flux focusing effect, concentrating magnetic field lines at chip edges and niobium film openings~\cite{Kapur2025, Shulga2025}. The corresponding magnetic field inhomogeneity can be on the order of a few tens of percent and is determined by a particular superconducting film layout, which makes it reproducible for a given design. Since a quantitative electromagnetic simulation of large and complex superconducting thin films remains challenging, we characterize this profile directly \textit{in situ}, using the signal phase accumulated along the SNAIL chain as a probe, and capture it with a minimal two-parameter model of the field distribution.

\begin{figure}[t]
    \centering
    \includegraphics[width=1.0\linewidth]{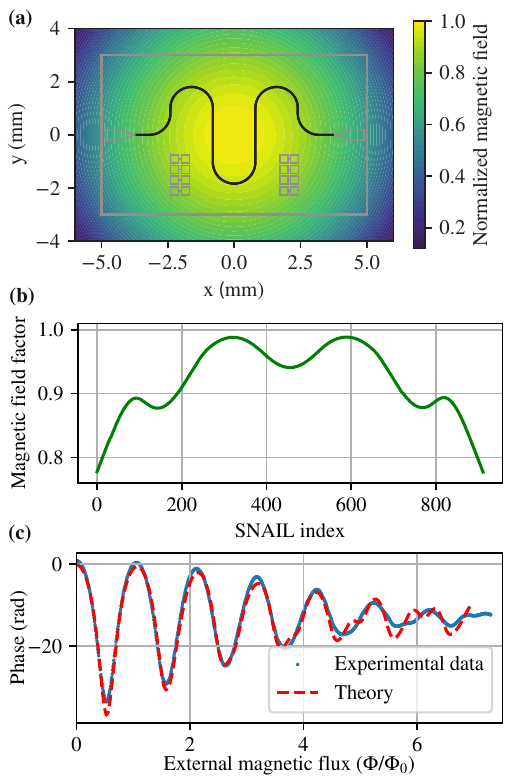}
    \caption{Effect of the magnetic field inhomogeneity across the JTWPA. (a) Normalized magnetic field distribution across the JTWPA chip (gray outline) according to Eq.\,\ref{eq:field_distribution}. The black line highlights the meandered SNAIL chain, which is used to estimate the resulting inhomogeneous magnetic field as shown in panel (b), where the magnetic field factor is the local flux normalized to its maximum value $\Phi_\mathrm{max}$ along the chain. Here, the edge SNAILs receive \qty{22}{\percent} less magnetic flux than the central ones, corresponding to $\Delta_0 = \num{0.15}$ at $(x_0,y_0) = \qty[parse-numbers = false]{(0,3)}{\milli\metre}$ in Eq.\,\ref{eq:field_distribution}. (c) Signal phase at the carrier frequency $f = \qty{4.0}{\giga\hertz}$, which is used to probe and extract the magnetic flux distribution. The best theory fit (red line) is obtained for $\Delta_0 = \num{0.15}$.
    }
    \label{fig:magnetic_field_distribution}
\end{figure}

Here, we use a simple model of magnetic field inhomogeneity by considering a two-dimensional Gaussian distribution with the maximum value of $\Phi_\mathrm{max}$:
\begin{equation}
    \Phi(x,y,\sigma) = \Phi_\mathrm{max}\exp{\left[-\frac{x^2+y^2}{2\sigma^2}\right]}\,,
    \label{eq:field_distribution}
\end{equation}
where we define the variance as $\sigma^2 =-\frac{x_0^2+y_0^2}{2 \ln(1-\Delta_0)} $, with $\Delta_0$ parameterizing the difference of magnetic flux between spatial points on the JTWPA chip with coordinates $(x_0, y_0)$ and $(0, 0)$. We plot the corresponding spatial field distribution in Fig.\,\ref{fig:magnetic_field_distribution}(a), while Fig.\,\ref{fig:magnetic_field_distribution}(b) shows the resulting flux dependence along the SNAIL line.

Experimentally, the magnetic field inhomogeneity can be probed by sending a microwave signal through the unpumped JTWPA at a fixed frequency and measuring its phase. An exemplary measurement trace is shown in Fig.\,\ref{fig:magnetic_field_distribution}(c). Here, we treat our system as a linear transmission line consisting of $N_\mathrm{cells}$ inductive elements whose values depend on the local magnetic field. Using the dimensionless dispersion relation derived from solving the linear wave equation of a transmission line made of the SNAIL inductances $L_\mathrm{J}$ with the corresponding total capacitances $C_\mathrm{J}$~\cite{Ranadive2022}
\begin{equation}
    k(\Phi) = \frac{\omega / \omega_0}{\sqrt{1-\omega^2/\omega_\mathrm{J}^2}}\,,
    \label{eq:dispersion}
\end{equation}
where $\omega_0 = 1/\sqrt{L_\mathrm{J}(\Phi)C_\mathrm{g}}$ is the cut-off frequency, and $\omega_\mathrm{J} = 1/\sqrt{L_\mathrm{J}(\Phi) C_\mathrm{J}}$ is the plasma frequency of the JTWPA, we can establish a connection between the experimentally measured phase dependence and the SNAIL inductance. By extension, the latter provides a link to the local magnetic field amplitude and allows us to fit the measured data with Eq.\,\ref{eq:dispersion}. In the case of an ideally homogeneous magnetic field, all SNAILs are biased with the same flux amplitude, so a corresponding phase drop over the whole chain can be derived from Eq.\,\ref{eq:dispersion} by multiplying it by $N_\mathrm{cells}$. In the case of an inhomogeneous magnetic field, each SNAIL contributes with a different phase drop to the propagating signal, since $L_\mathrm{J} = L^i_\mathrm{J}(\Phi(x,y))$ varies as a function of the local external magnetic field
\begin{equation}
    \theta(\Phi) = \theta_0 + \sum^{N_\mathrm{cells}}_{i=0}\frac{\omega  \sqrt{L_\mathrm{J}^i(\Phi) C_\mathrm{g}}}{\sqrt{1-\omega^2 L_\mathrm{J}^i(\Phi) C_\mathrm{J}}}\,,
    \label{eq:phase_inhomogeneous_field}
\end{equation}
where $\theta_0$ is a phase offset. Here, we integrate over the individual SNAIL phase drops in order to obtain the total accumulated signal phase, as shown in Fig.\,\ref{fig:magnetic_field_distribution}(c). A corresponding theory fit of the experimental data provides us with the inhomogeneity parameter $\Delta_0 = \num{0.15}$, which is substantially larger than the inhomogeneity value of around $\num{0.001}$ expected purely from the magnetic coil geometry, thus, confirming flux focusing in the superconducting film as the dominant contribution.

\section{Theory model}
\label{sec:model}

\begin{figure*}
    \centering
    \includegraphics[width=1.0\linewidth]{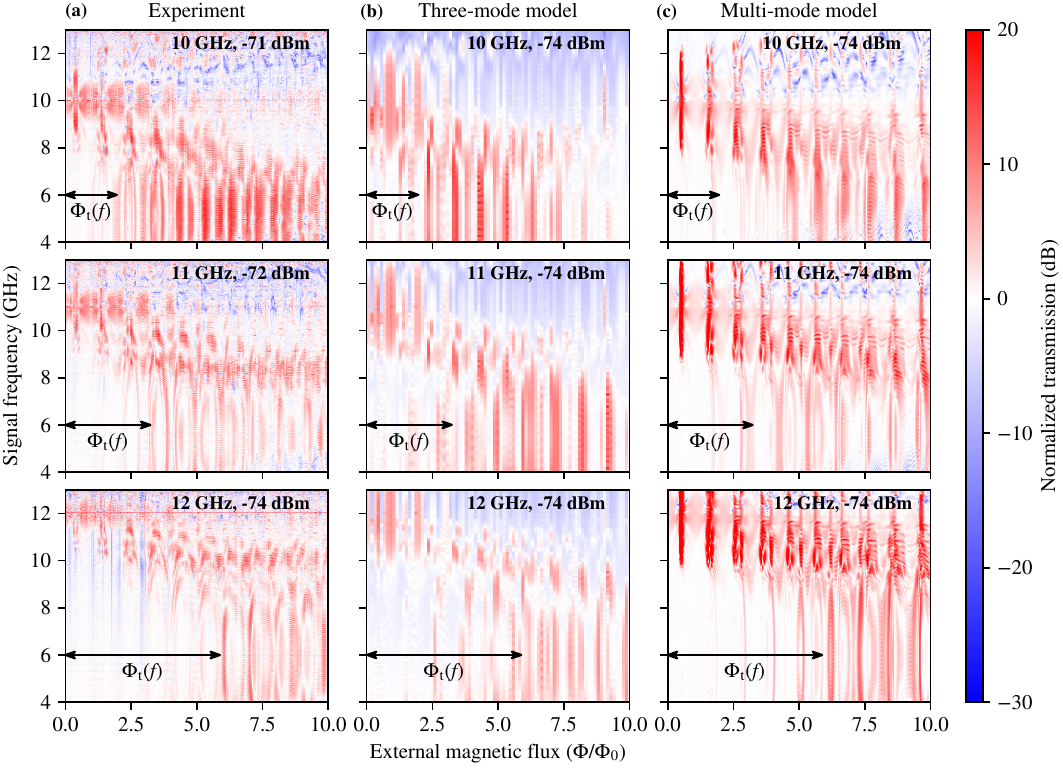}
    \caption{(a) Experimental normalized transmission measurements through the JTWPA as a function of the signal frequency and external magnetic flux for pump frequencies of \qtylist[list-units = single]{10;11;12}{\giga\hertz} at pump powers $P_\mathrm{p} =\qtylist[list-units = single]{-71;-72;-74}{\deci\belmilliwatt}$, correspondingly. At low values of the external magnetic flux $\Phi \sim \Phi_0$, the 4WM process is efficient due to a limited phase-matching. At larger values of $\Phi \gg \Phi_0$, the 3WM process becomes phase-matched, which leads to a broadband amplification response, characterized by a bandwidth $\sim \qty{4}{\giga\hertz}$. We mark corresponding threshold  values of magnetic flux with double arrows labeled $\Phi_\mathrm{t}(f)$. (b) Three-mode theoretical model of the JTWPA response in an inhomogeneous magnetic field at pump frequencies of \qtylist{10;11;12}{\giga\hertz} at a fixed pump power of $P_\mathrm{p} = \qty{-74}{\deci\belmilliwatt}$ (see Appendix \ref{app:three-mode-model} for more details). (c) Multi-mode harmonic-balance simulations of the JTWPA response in an inhomogeneous magnetic field for pump frequencies of \qtylist{10;11;12}{\giga\hertz} at a fixed pump power of $P_\mathrm{p} = \qty{-74}{\deci\belmilliwatt}$. Including additional mode mixing reveals higher-order effects beyond the three-mode approximation (see Appendix~\ref{app:multi-mode-model} for further details). The three-mode model evaluates the 3WM and 4WM gain separately and keeps the larger of the two at each probe frequency, whereas the multi-mode model solves both simultaneously and retains the additional mixing products.}
    \label{fig:exp_theory_comparison}
\end{figure*}

\begin{figure*}[t]
    \centering
    \includegraphics[width=1.0\linewidth]{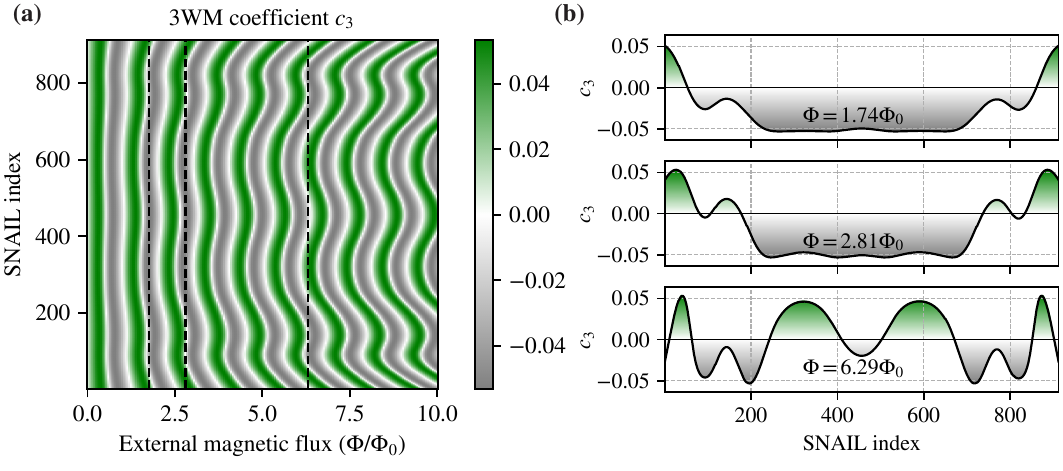}
    \caption{(a) Simulated nonlinearity $c_3$ of the SNAIL chain as a function of the inhomogeneous external magnetic flux based on Eq.\,\ref{eq:field_distribution}. Dashed black lines indicate threshold value $\Phi_\mathrm{t}(\{10, 11, 12\,\text{GHz}\}) = \{\num{1.74}, \num{2.81}, \num{6.29}\} \Phi_0$ corresponding to the emergence of the broadband 3WM parametric gain at the pump frequencies of \qtylist{10;11;12}{\giga\hertz} correspondingly,  as depicted in Fig.\,\ref{fig:exp_theory_comparison}.
    (b) Spatial profile of $c_3$ along the SNAIL chain at the three threshold flux values marked in dashed line on the plot on the left, showing how the domain structure becomes finer as the pump frequency increases.}
    \label{fig:domain_distribution}
\end{figure*}

In terms of the JTWPA physics, the magnetic field inhomogeneity leads to an inhomogeneous variation of nonlinear coefficients $c_3$ and $c_4$ across different SNAILs. For sufficiently large flux offsets, it may occur that the sign of $c_3$ changes across the SNAIL chain, as shown in Fig.\,\ref{fig:domain_distribution}(a). The $c_3$ sign reversal leads to an emerging phase-matching between the signal and pump modes~\cite{Armstrong1962, Zorin2021}. Here, the pump, signal, and idler modes are propagating through the JTWPA with different phase velocities and accumulate a respective phase difference $\Delta k = (k_\mathrm{p}-k_\mathrm{s}-k_\mathrm{i})$, where $k_{\mathrm{p},\mathrm{s},\mathrm{i}}$ are the dimensionless wave numbers of the pump, signal, and idler modes, respectively.
Large accumulated phase mismatches $\Delta k \sim \pi$ lead to a destructive interference between the modes and may completely suppress the parametric gain. However, in the regime with the periodic sign reversal, as described above, the phase mismatch does not become sufficiently large because the negative $c_3$ regions compensate for the positive ones, so that the resulting phase difference stays sufficiently small $(\Delta k \ll \pi)$, and the parametric amplification may reach reasonable gain values, before higher-order nonlinearities change this relatively simple picture. This intuitive model also explains different values of the threshold flux $\Phi_\mathrm{t}(f)$ as a function of the pump frequencies, as shown in Fig.\,\ref{fig:exp_theory_comparison}(b). Since lower pump frequencies exhibit lower phase mismatches it means that we need smaller negative $c_3$ areas and, respectively, smaller external magnetic flux offsets, to compensate for the arising phase mismatch. 

In order to validate this intuition and the measured dataset with a quantitative theoretical description, we develop two slightly different models: the first model is based on a  harmonic balance method with three coupled modes, while the second model takes into account multiple modes and their interference terms.

The three-mode model is based on the following approach:  first, we find minima of the individual SNAIL potentials and calculate corresponding equilibrium phases. Then, we expand the Josephson current–phase relation to the second and third orders to obtain the 3WM and 4WM nonlinear coefficients, $c_3$ and $c_4$, under the assumption of the inhomogeneous magnetic flux according to Eq.\,\ref{eq:field_distribution}. Next, we balance  harmonics of the nonlinear current along the SNAIL chain. This model retains only three modes: the pump, the signal, and the idler, neglecting higher intermixing products. The 3WM and 4WM gain values are evaluated separately, retaining the larger of the two at each probe frequency. More technical details about this model are provided in App.\,\ref{app:three-mode-model}. We plot the corresponding theory results in Fig.\,\ref{fig:exp_theory_comparison}(b) together with experimental observations provided in panel (a). As one can see, this rather simple model qualitatively reproduces the experimental results, which validates our hypothesis about the influence of the inhomogeneous magnetic flux on the JTWPA performance. 

The multi-mode model represents the harmonic response of each SNAIL unit cell by a Fourier series, retaining an arbitrary subset of generated mixing products, including a zero frequency component. Each SNAIL comprises symmetrically distributed capacitances to ground, an effective junction capacitance along the transmission line, and the nonlinear SNAIL element, together with relevant dielectric losses. The SNAIL nonlinearity is incorporated through a Taylor expansion of its current-phase relation about the minimum of its potential energy up to the cubic order in the phase excursion. This yields the inverse-inductance coefficient $c_2$, the 3WM coefficient $c_3$, and the 4WM coefficient $c_4$ for each individual unit cell. The phase corresponding to the minimum potential energy can be obtained analytically using, e.g. Cardano's formula after transforming the stationary condition $i(\varphi^\ast)=0$ into a cubic equation. This provides flux-bias-dependent, and consequently position-dependent, expressions for $c_2$, $c_3$, and $c_4$. These coefficients are incorporated into the individual unit-cell models according to the spatially varying flux bias of Eq.\,\ref{eq:field_distribution}.
The resulting chain of unit cells is terminated by ports that impose the boundary conditions at the input and output of the JTWPA. The unknown phase difference in each unit cell is represented by the Fourier coefficients of the retained harmonics. At the input, the externally applied signal and pump tones are specified at the signal and pump frequencies $\omega_\mathrm{s}$ and $\omega_\mathrm{p}$, respectively. Nonlinear mixing generates frequency components of the form $n_\mathrm{s}\omega_\mathrm{s}+n_\mathrm{p}\omega_\mathrm{p}$, where $n_\mathrm{s},n_\mathrm{p}\in\mathbb{Z}$.
The harmonic basis is truncated by imposing a maximum frequency, chosen, for example, in relation to the plasma frequency of the SNAIL unit cells, together with finite bounds on the integer indices $n_\mathrm{s}$ and $n_\mathrm{p}$. Within this basis, the 3WM processes are described by weighted convolutions of the Fourier coefficients, whereas the 4WM processes give rise to the corresponding repeated convolutions. Enforcing Kirchhoff's current law along the chain, together with the port boundary conditions, yields a sparse, multidimensional nonlinear system for the Fourier coefficients of the phase along the JTWPA. The different frequency components within each unit cell are densely coupled by the nonlinear mixing terms, while the spatial coupling remains sparse because each unit cell interacts only with its nearest neighbors.

The corresponding multi-mode simulation results are shown in the rightmost panels of Fig.\,\ref{fig:exp_theory_comparison}, together with the three-mode model and the experimental data. Both models reproduce the experimentally observed behavior and support the interpretation that a spatially inhomogeneous flux bias can enable gain through spatially engineered phase matching. More specifically, these models reproduce the existence of SNAIL domains with alternating $c_3$ sign and predict the matching threshold flux $\Phi_\mathrm{t}$ dependence on the pump frequency in comparison with the experimental data. For the magnetic field distribution used in the models, the number of sign reversals grows with the increase of applied flux, which shortens the domains length correspondingly (see Fig.~\ref{fig:domain_distribution}). The relevant comparison length is the coherence length $\ell_\mathrm{c}=\pi/\Delta k$\,\cite{Zorin2021}, which defines the number of cells after which the phase mismatch $\Delta k$, obtained from the dispersion relation of Eq.~\ref{eq:dispersion}, reaches the value $\pi$. At the 3WM sweet spot $\Phi = 0.354\,\Phi_0$ and for a degenerate signal, $\ell_\mathrm{c}$ amounts to 98, 70, and 50 cells for pump frequencies of \qtylist[list-units = single]{10;11;12}{\giga\hertz}. Therefore, a higher pump frequency requires a finer domain structure, which appears at higher values of magnetic flux. In Fig.~\ref{fig:domain_distribution}(b), we demonstrate the domain structure at flux values $\{\num{1.74}, \num{2.81}, \num{6.29}\} \Phi_0$ corresponding to threshold values of flux $\Phi_\mathrm{t}$ for pump frequencies of \qtylist[list-units = single]{10;11;12}{\giga\hertz}. Within each domain boundary the sign reversal compensates the accumulated phase mismatch, so the contribution of the domain adds to the broadband amplification rather than canceling.

\section{Conclusion and outlook}

We have investigated a Josephson traveling wave parametric amplifier  based on a homogeneous chain of $N_\mathrm{cells} = \num{912}$ superconducting nonlinear asymmetric inductive elements embedded in a coplanar waveguide geometry with shunting fishbone capacitors. This fully planar design has been fabricated using double-mask Manhattan-style overlap Josephson junctions which allow for the high junction yield and full flexibility in terms of in-plane rotations and area scaling. The corresponding JTWPA has been characterized in both the 3WM and 4WM regimes using conventional, vector network analyzer transmission measurements at temperatures of approximately \qty{15}{\milli\kelvin}. Here, we have uncovered a crossover from the 4WM-dominated broadband parametric amplification at small magnetic flux offsets $\Phi \sim \Phi_0$ to the 3WM-dominated regime at $\Phi \gg \Phi_0$. This behavior can be understood in terms of a field-induced phase-matching between signal, idler and parametric pump tones due to the spatial variation of the JTWPA nonlinear coefficients in an inhomogeneous magnetic field. The latter arises from the finite size of the superconducting coil providing the external magnetic flux bias to spatially distributed SNAILs as well as from flux-focusing effects in the superconducting CPW. 

More specifically, we have observed that the 4WM amplification process is getting progressively more suppressed as the external magnetic flux $\Phi$ increases, which can be understood in terms of increasingly nonuniform flux bias causing a greater disorder along the JTWPA. This flux variation reduces the uniformity of the 4WM response and leads to a suppression of the corresponding gain. In contrast, the 3WM process exhibits a markedly different and less intuitive dependence on the external magnetic field. At low magnetic fields, we have observed no measurable gain in this regime, while at higher fields,  starting from  $\Phi \gtrsim 6\Phi_0$ for the 12\,GHz parametric pump, a broadband amplification region with an approximate bandwidth $4-8$\,GHz emerges. Furthermore, the  onset of the 3WM amplification strongly depends on the pump frequency, with lower pump frequencies shifting the amplification region to lower threshold magnetic flux offsets.

We have interpreted these observations according to a phase-mismatch between the signal and pump modes due to the chromatic dispersion in the JTWPA. As the external magnetic field inhomogeneity increases, the accumulated flux variation causes some SNAILs to acquire an opposite sign of the cubic nonlinearity coefficient $c_3$, effectively reversing the sign of the local nonlinear interaction. This produces a phase-matching mechanism similar to quasi-phase matching in a poled nonlinear transmission line~\cite{Zorin2021}, enabling the broadband parametric gain at sufficiently high magnetic flux offset values $\Phi \gg \Phi_0$. The resulting amplification is sensitive to the spatial profile of the magnetic field distribution along the SNAIL array rather than to the average applied field amplitude. To test this interpretation, we have developed a quasi-classical model that quantitatively reproduces the experimental observations and supports the proposed field-induced, phase-matching mechanism. 

Ultimately, our experimental observations and numerical modeling results demonstrate that the magnetic field inhomogeneity, which is typically regarded as detrimental, can instead facilitate broadband parametric amplification in JTWPAs. We envision this approach as an interesting and viable alternative to the conventional phase-matching mechanisms, such as the resonant phase-matching. In particular, the absence of any resonances in the spectral bandwidth of parametric amplification reduces coupling to various thin-film impurities, which might result in a noticeable amplification noise reduction, as supported by added noise measurements with our JTWPA, see App.\,\ref{app:noise} for details. Purely from a fabrication perspective, it is much simpler to fabricate a fully homogeneous JTWPA, without any additional phase-matching elements, and target its desired parametric amplification properties by shaping a macroscopic profile of the SNAIL chain in the presence of a known magnetic field inhomogeneous profile. 

\begin{acknowledgments}
We acknowledge the support by the German Research Foundation via Germany's Excellence Strategy (EXC-2111-390814868) and by the German Federal Ministry of Education and Research via the project GeCQoS (FKZ 13N15680). This research is part of the Munich Quantum Valley, which is supported by the Bavarian state government with funds from the Hightech Agenda Bayern Plus. KS acknowledges the support by JSPS KAKENHI grant number 23H04864.
\end{acknowledgments}

\section{Data availability}

Experimental datasets are available upon a reasonable request.

\bibliography{refs}

\begin{thebibliography}{45}%
\makeatletter
\providecommand \@ifxundefined [1]{%
 \@ifx{#1\undefined}
}%
\providecommand \@ifnum [1]{%
 \ifnum #1\expandafter \@firstoftwo
 \else \expandafter \@secondoftwo
 \fi
}%
\providecommand \@ifx [1]{%
 \ifx #1\expandafter \@firstoftwo
 \else \expandafter \@secondoftwo
 \fi
}%
\providecommand \natexlab [1]{#1}%
\providecommand \enquote  [1]{``#1''}%
\providecommand \bibnamefont  [1]{#1}%
\providecommand \bibfnamefont [1]{#1}%
\providecommand \citenamefont [1]{#1}%
\providecommand \href@noop [0]{\@secondoftwo}%
\providecommand \href [0]{\begingroup \@sanitize@url \@href}%
\providecommand \@href[1]{\@@startlink{#1}\@@href}%
\providecommand \@@href[1]{\endgroup#1\@@endlink}%
\providecommand \@sanitize@url [0]{\catcode `\\12\catcode `\$12\catcode
  `\&12\catcode `\#12\catcode `\^12\catcode `\_12\catcode `\%12\relax}%
\providecommand \@@startlink[1]{}%
\providecommand \@@endlink[0]{}%
\providecommand \url  [0]{\begingroup\@sanitize@url \@url }%
\providecommand \@url [1]{\endgroup\@href {#1}{\urlprefix }}%
\providecommand \urlprefix  [0]{URL }%
\providecommand \Eprint [0]{\href }%
\providecommand \doibase [0]{https://doi.org/}%
\providecommand \selectlanguage [0]{\@gobble}%
\providecommand \bibinfo  [0]{\@secondoftwo}%
\providecommand \bibfield  [0]{\@secondoftwo}%
\providecommand \translation [1]{[#1]}%
\providecommand \BibitemOpen [0]{}%
\providecommand \bibitemStop [0]{}%
\providecommand \bibitemNoStop [0]{.\EOS\space}%
\providecommand \EOS [0]{\spacefactor3000\relax}%
\providecommand \BibitemShut  [1]{\csname bibitem#1\endcsname}%
\let\auto@bib@innerbib\@empty
\bibitem [{\citenamefont {Bartram}\ \emph {et~al.}(2023)\citenamefont
  {Bartram}, \citenamefont {Braine}, \citenamefont {Cervantes}, \citenamefont
  {Crisosto}, \citenamefont {Du}, \citenamefont {Leum}, \citenamefont
  {Mohapatra}, \citenamefont {Nitta}, \citenamefont {Rosenberg}, \citenamefont
  {Rybka}, \citenamefont {Yang}, \citenamefont {Clarke}, \citenamefont
  {Siddiqi}, \citenamefont {Agrawal}, \citenamefont {Dixit}, \citenamefont
  {Awida}, \citenamefont {Chou}, \citenamefont {Hollister}, \citenamefont
  {Knirck}, \citenamefont {Sonnenschein}, \citenamefont {Wester}, \citenamefont
  {Gleason}, \citenamefont {Hipp}, \citenamefont {Jois}, \citenamefont
  {Sikivie}, \citenamefont {Sullivan}, \citenamefont {Tanner}, \citenamefont
  {Lentz}, \citenamefont {Khatiwada}, \citenamefont {Carosi}, \citenamefont
  {Cisneros}, \citenamefont {Robertson}, \citenamefont {Woollett},
  \citenamefont {Duffy}, \citenamefont {Boutan}, \citenamefont {Jones},
  \citenamefont {LaRoque}, \citenamefont {Oblath}, \citenamefont {Taubman},
  \citenamefont {Daw}, \citenamefont {Perry}, \citenamefont {Buckley},
  \citenamefont {Gaikwad}, \citenamefont {Hoffman}, \citenamefont {Murch},
  \citenamefont {Goryachev}, \citenamefont {McAllister}, \citenamefont
  {Quiskamp}, \citenamefont {Thomson}, \citenamefont {Tobar}, \citenamefont
  {Bolkhovsky}, \citenamefont {Calusine}, \citenamefont {Oliver},\ and\
  \citenamefont {Serniak}}]{Bartram2023}%
  \BibitemOpen
  \bibfield  {author} {\bibinfo {author} {\bibfnamefont {C.}~\bibnamefont
  {Bartram}}, \bibinfo {author} {\bibfnamefont {T.}~\bibnamefont {Braine}},
  \bibinfo {author} {\bibfnamefont {R.}~\bibnamefont {Cervantes}}, \bibinfo
  {author} {\bibfnamefont {N.}~\bibnamefont {Crisosto}}, \bibinfo {author}
  {\bibfnamefont {N.}~\bibnamefont {Du}}, \bibinfo {author} {\bibfnamefont
  {G.}~\bibnamefont {Leum}}, \bibinfo {author} {\bibfnamefont {P.}~\bibnamefont
  {Mohapatra}}, \bibinfo {author} {\bibfnamefont {T.}~\bibnamefont {Nitta}},
  \bibinfo {author} {\bibfnamefont {L.~J.}\ \bibnamefont {Rosenberg}}, \bibinfo
  {author} {\bibfnamefont {G.}~\bibnamefont {Rybka}}, \bibinfo {author}
  {\bibfnamefont {J.}~\bibnamefont {Yang}}, \bibinfo {author} {\bibfnamefont
  {J.}~\bibnamefont {Clarke}}, \bibinfo {author} {\bibfnamefont
  {I.}~\bibnamefont {Siddiqi}}, \bibinfo {author} {\bibfnamefont
  {A.}~\bibnamefont {Agrawal}}, \bibinfo {author} {\bibfnamefont {A.~V.}\
  \bibnamefont {Dixit}}, \bibinfo {author} {\bibfnamefont {M.~H.}\ \bibnamefont
  {Awida}}, \bibinfo {author} {\bibfnamefont {A.~S.}\ \bibnamefont {Chou}},
  \bibinfo {author} {\bibfnamefont {M.}~\bibnamefont {Hollister}}, \bibinfo
  {author} {\bibfnamefont {S.}~\bibnamefont {Knirck}}, \bibinfo {author}
  {\bibfnamefont {A.}~\bibnamefont {Sonnenschein}}, \bibinfo {author}
  {\bibfnamefont {W.}~\bibnamefont {Wester}}, \bibinfo {author} {\bibfnamefont
  {J.~R.}\ \bibnamefont {Gleason}}, \bibinfo {author} {\bibfnamefont {A.~T.}\
  \bibnamefont {Hipp}}, \bibinfo {author} {\bibfnamefont {S.}~\bibnamefont
  {Jois}}, \bibinfo {author} {\bibfnamefont {P.}~\bibnamefont {Sikivie}},
  \bibinfo {author} {\bibfnamefont {N.~S.}\ \bibnamefont {Sullivan}}, \bibinfo
  {author} {\bibfnamefont {D.~B.}\ \bibnamefont {Tanner}}, \bibinfo {author}
  {\bibfnamefont {E.}~\bibnamefont {Lentz}}, \bibinfo {author} {\bibfnamefont
  {R.}~\bibnamefont {Khatiwada}}, \bibinfo {author} {\bibfnamefont
  {G.}~\bibnamefont {Carosi}}, \bibinfo {author} {\bibfnamefont
  {C.}~\bibnamefont {Cisneros}}, \bibinfo {author} {\bibfnamefont
  {N.}~\bibnamefont {Robertson}}, \bibinfo {author} {\bibfnamefont
  {N.}~\bibnamefont {Woollett}}, \bibinfo {author} {\bibfnamefont {L.~D.}\
  \bibnamefont {Duffy}}, \bibinfo {author} {\bibfnamefont {C.}~\bibnamefont
  {Boutan}}, \bibinfo {author} {\bibfnamefont {M.}~\bibnamefont {Jones}},
  \bibinfo {author} {\bibfnamefont {B.~H.}\ \bibnamefont {LaRoque}}, \bibinfo
  {author} {\bibfnamefont {N.~S.}\ \bibnamefont {Oblath}}, \bibinfo {author}
  {\bibfnamefont {M.~S.}\ \bibnamefont {Taubman}}, \bibinfo {author}
  {\bibfnamefont {E.~J.}\ \bibnamefont {Daw}}, \bibinfo {author} {\bibfnamefont
  {M.~G.}\ \bibnamefont {Perry}}, \bibinfo {author} {\bibfnamefont {J.~H.}\
  \bibnamefont {Buckley}}, \bibinfo {author} {\bibfnamefont {C.}~\bibnamefont
  {Gaikwad}}, \bibinfo {author} {\bibfnamefont {J.}~\bibnamefont {Hoffman}},
  \bibinfo {author} {\bibfnamefont {K.}~\bibnamefont {Murch}}, \bibinfo
  {author} {\bibfnamefont {M.}~\bibnamefont {Goryachev}}, \bibinfo {author}
  {\bibfnamefont {B.~T.}\ \bibnamefont {McAllister}}, \bibinfo {author}
  {\bibfnamefont {A.}~\bibnamefont {Quiskamp}}, \bibinfo {author}
  {\bibfnamefont {C.}~\bibnamefont {Thomson}}, \bibinfo {author} {\bibfnamefont
  {M.~E.}\ \bibnamefont {Tobar}}, \bibinfo {author} {\bibfnamefont
  {V.}~\bibnamefont {Bolkhovsky}}, \bibinfo {author} {\bibfnamefont
  {G.}~\bibnamefont {Calusine}}, \bibinfo {author} {\bibfnamefont
  {W.}~\bibnamefont {Oliver}},\ and\ \bibinfo {author} {\bibfnamefont
  {K.}~\bibnamefont {Serniak}},\ }\bibfield  {title} {\bibinfo {title} {Dark
  matter axion search using a {J}osephson traveling wave parametric
  amplifier},\ }\href {https://doi.org/10.1063/5.0122907} {\bibfield  {journal}
  {\bibinfo  {journal} {Rev. Sci. Instrum.}\ }\textbf {\bibinfo {volume}
  {94}},\ \bibinfo {pages} {044703} (\bibinfo {year} {2023})}\BibitemShut
  {NoStop}%
\bibitem [{\citenamefont {Bravyi}\ \emph {et~al.}(2022)\citenamefont {Bravyi},
  \citenamefont {Dial}, \citenamefont {Gambetta}, \citenamefont {Gil},\ and\
  \citenamefont {Nazario}}]{Bravyi2022}%
  \BibitemOpen
  \bibfield  {author} {\bibinfo {author} {\bibfnamefont {S.}~\bibnamefont
  {Bravyi}}, \bibinfo {author} {\bibfnamefont {O.}~\bibnamefont {Dial}},
  \bibinfo {author} {\bibfnamefont {J.~M.}\ \bibnamefont {Gambetta}}, \bibinfo
  {author} {\bibfnamefont {D.}~\bibnamefont {Gil}},\ and\ \bibinfo {author}
  {\bibfnamefont {Z.}~\bibnamefont {Nazario}},\ }\bibfield  {title} {\bibinfo
  {title} {The future of quantum computing with superconducting qubits},\
  }\href {https://doi.org/10.1063/5.0082975} {\bibfield  {journal} {\bibinfo
  {journal} {J. Appl. Phys.}\ }\textbf {\bibinfo {volume} {132}},\ \bibinfo
  {pages} {160902} (\bibinfo {year} {2022})}\BibitemShut {NoStop}%
\bibitem [{\citenamefont {Mohseni}\ \emph {et~al.}(2024)\citenamefont
  {Mohseni}, \citenamefont {Scherer}, \citenamefont {Johnson}, \citenamefont
  {Wertheim}, \citenamefont {Otten}, \citenamefont {Aadit}, \citenamefont
  {Alexeev}, \citenamefont {Bresniker}, \citenamefont {Camsari}, \citenamefont
  {Chapman}, \citenamefont {Chatterjee}, \citenamefont {Dagnew}, \citenamefont
  {Esposito}, \citenamefont {Fahim}, \citenamefont {Fiorentino}, \citenamefont
  {Gajjar}, \citenamefont {Khalid}, \citenamefont {Kong}, \citenamefont
  {Kulchytskyy}, \citenamefont {Kyoseva}, \citenamefont {Li}, \citenamefont
  {Lott}, \citenamefont {Markov}, \citenamefont {McDermott}, \citenamefont
  {Pedretti}, \citenamefont {Rao}, \citenamefont {Rieffel}, \citenamefont
  {Silva}, \citenamefont {Sorebo}, \citenamefont {Spentzouris}, \citenamefont
  {Steiner}, \citenamefont {Torosov}, \citenamefont {Venturelli}, \citenamefont
  {Visser}, \citenamefont {Webb}, \citenamefont {Zhan}, \citenamefont {Cohen},
  \citenamefont {Ronagh}, \citenamefont {Ho}, \citenamefont {Beausoleil},\ and\
  \citenamefont {Martinis}}]{Martinis2025}%
  \BibitemOpen
  \bibfield  {author} {\bibinfo {author} {\bibfnamefont {M.}~\bibnamefont
  {Mohseni}}, \bibinfo {author} {\bibfnamefont {A.}~\bibnamefont {Scherer}},
  \bibinfo {author} {\bibfnamefont {K.~G.}\ \bibnamefont {Johnson}}, \bibinfo
  {author} {\bibfnamefont {O.}~\bibnamefont {Wertheim}}, \bibinfo {author}
  {\bibfnamefont {M.}~\bibnamefont {Otten}}, \bibinfo {author} {\bibfnamefont
  {N.~A.}\ \bibnamefont {Aadit}}, \bibinfo {author} {\bibfnamefont
  {Y.}~\bibnamefont {Alexeev}}, \bibinfo {author} {\bibfnamefont {K.~M.}\
  \bibnamefont {Bresniker}}, \bibinfo {author} {\bibfnamefont {K.~Y.}\
  \bibnamefont {Camsari}}, \bibinfo {author} {\bibfnamefont {B.}~\bibnamefont
  {Chapman}}, \bibinfo {author} {\bibfnamefont {S.}~\bibnamefont {Chatterjee}},
  \bibinfo {author} {\bibfnamefont {G.~A.}\ \bibnamefont {Dagnew}}, \bibinfo
  {author} {\bibfnamefont {A.}~\bibnamefont {Esposito}}, \bibinfo {author}
  {\bibfnamefont {F.}~\bibnamefont {Fahim}}, \bibinfo {author} {\bibfnamefont
  {M.}~\bibnamefont {Fiorentino}}, \bibinfo {author} {\bibfnamefont
  {A.}~\bibnamefont {Gajjar}}, \bibinfo {author} {\bibfnamefont
  {A.}~\bibnamefont {Khalid}}, \bibinfo {author} {\bibfnamefont
  {X.}~\bibnamefont {Kong}}, \bibinfo {author} {\bibfnamefont {B.}~\bibnamefont
  {Kulchytskyy}}, \bibinfo {author} {\bibfnamefont {E.}~\bibnamefont
  {Kyoseva}}, \bibinfo {author} {\bibfnamefont {R.}~\bibnamefont {Li}},
  \bibinfo {author} {\bibfnamefont {P.~A.}\ \bibnamefont {Lott}}, \bibinfo
  {author} {\bibfnamefont {I.~L.}\ \bibnamefont {Markov}}, \bibinfo {author}
  {\bibfnamefont {R.~F.}\ \bibnamefont {McDermott}}, \bibinfo {author}
  {\bibfnamefont {G.}~\bibnamefont {Pedretti}}, \bibinfo {author}
  {\bibfnamefont {P.}~\bibnamefont {Rao}}, \bibinfo {author} {\bibfnamefont
  {E.}~\bibnamefont {Rieffel}}, \bibinfo {author} {\bibfnamefont
  {A.}~\bibnamefont {Silva}}, \bibinfo {author} {\bibfnamefont
  {J.}~\bibnamefont {Sorebo}}, \bibinfo {author} {\bibfnamefont
  {P.}~\bibnamefont {Spentzouris}}, \bibinfo {author} {\bibfnamefont
  {Z.}~\bibnamefont {Steiner}}, \bibinfo {author} {\bibfnamefont
  {B.}~\bibnamefont {Torosov}}, \bibinfo {author} {\bibfnamefont
  {D.}~\bibnamefont {Venturelli}}, \bibinfo {author} {\bibfnamefont {R.~J.}\
  \bibnamefont {Visser}}, \bibinfo {author} {\bibfnamefont {Z.}~\bibnamefont
  {Webb}}, \bibinfo {author} {\bibfnamefont {X.}~\bibnamefont {Zhan}}, \bibinfo
  {author} {\bibfnamefont {Y.}~\bibnamefont {Cohen}}, \bibinfo {author}
  {\bibfnamefont {P.}~\bibnamefont {Ronagh}}, \bibinfo {author} {\bibfnamefont
  {A.}~\bibnamefont {Ho}}, \bibinfo {author} {\bibfnamefont {R.~G.}\
  \bibnamefont {Beausoleil}},\ and\ \bibinfo {author} {\bibfnamefont {J.~M.}\
  \bibnamefont {Martinis}},\ }\href@noop {} {\bibinfo {title} {How to build a
  quantum supercomputer: {S}caling from hundreds to millions of qubits}}
  (\bibinfo {year} {2024}),\ \Eprint {https://arxiv.org/abs/2411.10406}
  {arXiv:2411.10406 [quant-ph]} \BibitemShut {NoStop}%
\bibitem [{\citenamefont {Esposito}\ \emph {et~al.}(2022)\citenamefont
  {Esposito}, \citenamefont {Ranadive}, \citenamefont {Planat}, \citenamefont
  {L\'eger}, \citenamefont {Fraudet}, \citenamefont {Jouanny}, \citenamefont
  {Buisson}, \citenamefont {Guichard}, \citenamefont {Naud}, \citenamefont
  {Aumentado}, \citenamefont {Lecocq},\ and\ \citenamefont
  {Roch}}]{Esposito2022}%
  \BibitemOpen
  \bibfield  {author} {\bibinfo {author} {\bibfnamefont {M.}~\bibnamefont
  {Esposito}}, \bibinfo {author} {\bibfnamefont {A.}~\bibnamefont {Ranadive}},
  \bibinfo {author} {\bibfnamefont {L.}~\bibnamefont {Planat}}, \bibinfo
  {author} {\bibfnamefont {S.}~\bibnamefont {L\'eger}}, \bibinfo {author}
  {\bibfnamefont {D.}~\bibnamefont {Fraudet}}, \bibinfo {author} {\bibfnamefont
  {V.}~\bibnamefont {Jouanny}}, \bibinfo {author} {\bibfnamefont
  {O.}~\bibnamefont {Buisson}}, \bibinfo {author} {\bibfnamefont
  {W.}~\bibnamefont {Guichard}}, \bibinfo {author} {\bibfnamefont
  {C.}~\bibnamefont {Naud}}, \bibinfo {author} {\bibfnamefont {J.}~\bibnamefont
  {Aumentado}}, \bibinfo {author} {\bibfnamefont {F.}~\bibnamefont {Lecocq}},\
  and\ \bibinfo {author} {\bibfnamefont {N.}~\bibnamefont {Roch}},\ }\bibfield
  {title} {\bibinfo {title} {Observation of two-mode squeezing in a traveling
  wave parametric amplifier},\ }\href
  {https://doi.org/10.1103/PhysRevLett.128.153603} {\bibfield  {journal}
  {\bibinfo  {journal} {Phys. Rev. Lett.}\ }\textbf {\bibinfo {volume} {128}},\
  \bibinfo {pages} {153603} (\bibinfo {year} {2022})}\BibitemShut {NoStop}%
\bibitem [{\citenamefont {Qiu}\ \emph {et~al.}(2023)\citenamefont {Qiu},
  \citenamefont {Grimsmo}, \citenamefont {Peng}, \citenamefont {Kannan},
  \citenamefont {Lienhard}, \citenamefont {Sung}, \citenamefont {Krantz},
  \citenamefont {Bolkhovsky}, \citenamefont {Calusine}, \citenamefont {Kim},
  \citenamefont {Melville}, \citenamefont {Niedzielski}, \citenamefont {Yoder},
  \citenamefont {Schwartz}, \citenamefont {Orlando}, \citenamefont {Siddiqi},
  \citenamefont {Gustavsson}, \citenamefont {O'Brien},\ and\ \citenamefont
  {Oliver}}]{Qiu2023}%
  \BibitemOpen
  \bibfield  {author} {\bibinfo {author} {\bibfnamefont {J.~Y.}\ \bibnamefont
  {Qiu}}, \bibinfo {author} {\bibfnamefont {A.}~\bibnamefont {Grimsmo}},
  \bibinfo {author} {\bibfnamefont {K.}~\bibnamefont {Peng}}, \bibinfo {author}
  {\bibfnamefont {B.}~\bibnamefont {Kannan}}, \bibinfo {author} {\bibfnamefont
  {B.}~\bibnamefont {Lienhard}}, \bibinfo {author} {\bibfnamefont
  {Y.}~\bibnamefont {Sung}}, \bibinfo {author} {\bibfnamefont {P.}~\bibnamefont
  {Krantz}}, \bibinfo {author} {\bibfnamefont {V.}~\bibnamefont {Bolkhovsky}},
  \bibinfo {author} {\bibfnamefont {G.}~\bibnamefont {Calusine}}, \bibinfo
  {author} {\bibfnamefont {D.}~\bibnamefont {Kim}}, \bibinfo {author}
  {\bibfnamefont {A.}~\bibnamefont {Melville}}, \bibinfo {author}
  {\bibfnamefont {B.~M.}\ \bibnamefont {Niedzielski}}, \bibinfo {author}
  {\bibfnamefont {J.}~\bibnamefont {Yoder}}, \bibinfo {author} {\bibfnamefont
  {M.~E.}\ \bibnamefont {Schwartz}}, \bibinfo {author} {\bibfnamefont {T.~P.}\
  \bibnamefont {Orlando}}, \bibinfo {author} {\bibfnamefont {I.}~\bibnamefont
  {Siddiqi}}, \bibinfo {author} {\bibfnamefont {S.}~\bibnamefont {Gustavsson}},
  \bibinfo {author} {\bibfnamefont {K.~P.}\ \bibnamefont {O'Brien}},\ and\
  \bibinfo {author} {\bibfnamefont {W.~D.}\ \bibnamefont {Oliver}},\ }\bibfield
   {title} {\bibinfo {title} {Broadband squeezed microwaves and amplification
  with a {J}osephson travelling-wave parametric amplifier},\ }\href
  {https://doi.org/10.1038/s41567-022-01929-w} {\bibfield  {journal} {\bibinfo
  {journal} {Nat. Phys.}\ }\textbf {\bibinfo {volume} {19}},\ \bibinfo {pages}
  {706} (\bibinfo {year} {2023})}\BibitemShut {NoStop}%
\bibitem [{\citenamefont {Pfister}(2020)}]{Pfister2020}%
  \BibitemOpen
  \bibfield  {author} {\bibinfo {author} {\bibfnamefont {O.}~\bibnamefont
  {Pfister}},\ }\bibfield  {title} {\bibinfo {title} {Continuous-variable
  quantum computing in the quantum optical frequency comb},\ }\href
  {https://doi.org/10.1088/1361-6455/ab526f} {\bibfield  {journal} {\bibinfo
  {journal} {J. Phys. B: At. Mol. Opt. Phys.}\ }\textbf {\bibinfo {volume}
  {53}},\ \bibinfo {pages} {012001} (\bibinfo {year} {2020})}\BibitemShut
  {NoStop}%
\bibitem [{\citenamefont {Lingua}\ \emph {et~al.}(2025)\citenamefont {Lingua},
  \citenamefont {Rivera~Hern{\'a}ndez}, \citenamefont {Cortinovis},\ and\
  \citenamefont {Haviland}}]{Lingua2025}%
  \BibitemOpen
  \bibfield  {author} {\bibinfo {author} {\bibfnamefont {F.}~\bibnamefont
  {Lingua}}, \bibinfo {author} {\bibfnamefont {J.~C.}\ \bibnamefont
  {Rivera~Hern{\'a}ndez}}, \bibinfo {author} {\bibfnamefont {M.}~\bibnamefont
  {Cortinovis}},\ and\ \bibinfo {author} {\bibfnamefont {D.~B.}\ \bibnamefont
  {Haviland}},\ }\bibfield  {title} {\bibinfo {title} {Continuous-variable
  square-ladder cluster states in a microwave frequency comb},\ }\href
  {https://doi.org/10.1103/PhysRevLett.134.183602} {\bibfield  {journal}
  {\bibinfo  {journal} {Phys. Rev. Lett.}\ }\textbf {\bibinfo {volume} {134}},\
  \bibinfo {pages} {183602} (\bibinfo {year} {2025})}\BibitemShut {NoStop}%
\bibitem [{\citenamefont {Fedorov}\ \emph {et~al.}(2021)\citenamefont
  {Fedorov}, \citenamefont {Renger}, \citenamefont {Pogorzalek}, \citenamefont
  {Di~Candia}, \citenamefont {Chen}, \citenamefont {Nojiri}, \citenamefont
  {Inomata}, \citenamefont {Nakamura}, \citenamefont {Partanen}, \citenamefont
  {Marx}, \citenamefont {Gross},\ and\ \citenamefont {Deppe}}]{Fedorov2021}%
  \BibitemOpen
  \bibfield  {author} {\bibinfo {author} {\bibfnamefont {K.~G.}\ \bibnamefont
  {Fedorov}}, \bibinfo {author} {\bibfnamefont {M.}~\bibnamefont {Renger}},
  \bibinfo {author} {\bibfnamefont {S.}~\bibnamefont {Pogorzalek}}, \bibinfo
  {author} {\bibfnamefont {R.}~\bibnamefont {Di~Candia}}, \bibinfo {author}
  {\bibfnamefont {Q.}~\bibnamefont {Chen}}, \bibinfo {author} {\bibfnamefont
  {Y.}~\bibnamefont {Nojiri}}, \bibinfo {author} {\bibfnamefont
  {K.}~\bibnamefont {Inomata}}, \bibinfo {author} {\bibfnamefont
  {Y.}~\bibnamefont {Nakamura}}, \bibinfo {author} {\bibfnamefont
  {M.}~\bibnamefont {Partanen}}, \bibinfo {author} {\bibfnamefont
  {A.}~\bibnamefont {Marx}}, \bibinfo {author} {\bibfnamefont {R.}~\bibnamefont
  {Gross}},\ and\ \bibinfo {author} {\bibfnamefont {F.}~\bibnamefont {Deppe}},\
  }\bibfield  {title} {\bibinfo {title} {Experimental quantum teleportation of
  propagating microwaves},\ }\href {https://doi.org/10.1126/sciadv.abk0891}
  {\bibfield  {journal} {\bibinfo  {journal} {Sci. Adv.}\ }\textbf {\bibinfo
  {volume} {7}},\ \bibinfo {pages} {eabk0891} (\bibinfo {year}
  {2021})}\BibitemShut {NoStop}%
\bibitem [{\citenamefont {Fesquet}\ \emph {et~al.}(2023)\citenamefont
  {Fesquet}, \citenamefont {Kronowetter}, \citenamefont {Renger}, \citenamefont
  {Chen}, \citenamefont {Honasoge}, \citenamefont {Gargiulo}, \citenamefont
  {Nojiri}, \citenamefont {Marx}, \citenamefont {Deppe}, \citenamefont
  {Gross},\ and\ \citenamefont {Fedorov}}]{Fesquet2023}%
  \BibitemOpen
  \bibfield  {author} {\bibinfo {author} {\bibfnamefont {F.}~\bibnamefont
  {Fesquet}}, \bibinfo {author} {\bibfnamefont {F.}~\bibnamefont
  {Kronowetter}}, \bibinfo {author} {\bibfnamefont {M.}~\bibnamefont {Renger}},
  \bibinfo {author} {\bibfnamefont {Q.}~\bibnamefont {Chen}}, \bibinfo {author}
  {\bibfnamefont {K.}~\bibnamefont {Honasoge}}, \bibinfo {author}
  {\bibfnamefont {O.}~\bibnamefont {Gargiulo}}, \bibinfo {author}
  {\bibfnamefont {Y.}~\bibnamefont {Nojiri}}, \bibinfo {author} {\bibfnamefont
  {A.}~\bibnamefont {Marx}}, \bibinfo {author} {\bibfnamefont {F.}~\bibnamefont
  {Deppe}}, \bibinfo {author} {\bibfnamefont {R.}~\bibnamefont {Gross}},\ and\
  \bibinfo {author} {\bibfnamefont {K.~G.}\ \bibnamefont {Fedorov}},\
  }\bibfield  {title} {\bibinfo {title} {Perspectives of microwave quantum key
  distribution in the open air},\ }\href
  {https://doi.org/10.1103/PhysRevA.108.032607} {\bibfield  {journal} {\bibinfo
   {journal} {Phys. Rev. A}\ }\textbf {\bibinfo {volume} {108}},\ \bibinfo
  {pages} {032607} (\bibinfo {year} {2023})}\BibitemShut {NoStop}%
\bibitem [{\citenamefont {Frattini}\ \emph {et~al.}(2017)\citenamefont
  {Frattini}, \citenamefont {Vool}, \citenamefont {Shankar}, \citenamefont
  {Narla}, \citenamefont {Sliwa},\ and\ \citenamefont
  {Devoret}}]{Frattini2017}%
  \BibitemOpen
  \bibfield  {author} {\bibinfo {author} {\bibfnamefont {N.~E.}\ \bibnamefont
  {Frattini}}, \bibinfo {author} {\bibfnamefont {U.}~\bibnamefont {Vool}},
  \bibinfo {author} {\bibfnamefont {S.}~\bibnamefont {Shankar}}, \bibinfo
  {author} {\bibfnamefont {A.}~\bibnamefont {Narla}}, \bibinfo {author}
  {\bibfnamefont {K.~M.}\ \bibnamefont {Sliwa}},\ and\ \bibinfo {author}
  {\bibfnamefont {M.~H.}\ \bibnamefont {Devoret}},\ }\bibfield  {title}
  {\bibinfo {title} {3-wave mixing {J}osephson dipole element},\ }\href
  {https://doi.org/10.1063/1.4984142} {\bibfield  {journal} {\bibinfo
  {journal} {Appl. Phys. Lett.}\ }\textbf {\bibinfo {volume} {110}},\ \bibinfo
  {pages} {222603} (\bibinfo {year} {2017})}\BibitemShut {NoStop}%
\bibitem [{\citenamefont {O’Brien}\ \emph {et~al.}(2014)\citenamefont
  {O’Brien}, \citenamefont {Macklin}, \citenamefont {Siddiqi},\ and\
  \citenamefont {Zhang}}]{O’Brien2014}%
  \BibitemOpen
  \bibfield  {author} {\bibinfo {author} {\bibfnamefont {K.}~\bibnamefont
  {O’Brien}}, \bibinfo {author} {\bibfnamefont {C.}~\bibnamefont {Macklin}},
  \bibinfo {author} {\bibfnamefont {I.}~\bibnamefont {Siddiqi}},\ and\ \bibinfo
  {author} {\bibfnamefont {X.}~\bibnamefont {Zhang}},\ }\bibfield  {title}
  {\bibinfo {title} {Resonant phase matching of {Josephson} junction traveling
  wave parametric amplifiers},\ }\href
  {https://doi.org/10.1103/physrevlett.113.157001} {\bibfield  {journal}
  {\bibinfo  {journal} {Phys. Rev. Lett.}\ }\textbf {\bibinfo {volume} {113}},\
  \bibinfo {pages} {157001} (\bibinfo {year} {2014})}\BibitemShut {NoStop}%
\bibitem [{\citenamefont {Macklin}\ \emph {et~al.}(2015)\citenamefont
  {Macklin}, \citenamefont {O'Brien}, \citenamefont {Hover}, \citenamefont
  {Schwartz}, \citenamefont {Bolkhovsky}, \citenamefont {Zhang}, \citenamefont
  {Oliver},\ and\ \citenamefont {Siddiqi}}]{Macklin2015}%
  \BibitemOpen
  \bibfield  {author} {\bibinfo {author} {\bibfnamefont {C.}~\bibnamefont
  {Macklin}}, \bibinfo {author} {\bibfnamefont {K.}~\bibnamefont {O'Brien}},
  \bibinfo {author} {\bibfnamefont {D.}~\bibnamefont {Hover}}, \bibinfo
  {author} {\bibfnamefont {M.~E.}\ \bibnamefont {Schwartz}}, \bibinfo {author}
  {\bibfnamefont {V.}~\bibnamefont {Bolkhovsky}}, \bibinfo {author}
  {\bibfnamefont {X.}~\bibnamefont {Zhang}}, \bibinfo {author} {\bibfnamefont
  {W.~D.}\ \bibnamefont {Oliver}},\ and\ \bibinfo {author} {\bibfnamefont
  {I.}~\bibnamefont {Siddiqi}},\ }\bibfield  {title} {\bibinfo {title} {A
  near-quantum-limited {J}osephson traveling-wave parametric amplifier},\
  }\href {https://doi.org/10.1126/science.aaa8525} {\bibfield  {journal}
  {\bibinfo  {journal} {Science}\ }\textbf {\bibinfo {volume} {350}},\ \bibinfo
  {pages} {307} (\bibinfo {year} {2015})}\BibitemShut {NoStop}%
\bibitem [{\citenamefont {White}\ \emph {et~al.}(2015)\citenamefont {White},
  \citenamefont {Mutus}, \citenamefont {Hoi}, \citenamefont {Barends},
  \citenamefont {Campbell}, \citenamefont {Chen}, \citenamefont {Chen},
  \citenamefont {Chiaro}, \citenamefont {Dunsworth}, \citenamefont {Jeffrey},
  \citenamefont {Kelly}, \citenamefont {Megrant}, \citenamefont {Neill},
  \citenamefont {O'Malley}, \citenamefont {Roushan}, \citenamefont {Sank},
  \citenamefont {Vainsencher}, \citenamefont {Wenner}, \citenamefont
  {Chaudhuri}, \citenamefont {Gao},\ and\ \citenamefont
  {Martinis}}]{White2015}%
  \BibitemOpen
  \bibfield  {author} {\bibinfo {author} {\bibfnamefont {T.~C.}\ \bibnamefont
  {White}}, \bibinfo {author} {\bibfnamefont {J.~Y.}\ \bibnamefont {Mutus}},
  \bibinfo {author} {\bibfnamefont {I.-C.}\ \bibnamefont {Hoi}}, \bibinfo
  {author} {\bibfnamefont {R.}~\bibnamefont {Barends}}, \bibinfo {author}
  {\bibfnamefont {B.}~\bibnamefont {Campbell}}, \bibinfo {author}
  {\bibfnamefont {Y.}~\bibnamefont {Chen}}, \bibinfo {author} {\bibfnamefont
  {Z.}~\bibnamefont {Chen}}, \bibinfo {author} {\bibfnamefont {B.}~\bibnamefont
  {Chiaro}}, \bibinfo {author} {\bibfnamefont {A.}~\bibnamefont {Dunsworth}},
  \bibinfo {author} {\bibfnamefont {E.}~\bibnamefont {Jeffrey}}, \bibinfo
  {author} {\bibfnamefont {J.}~\bibnamefont {Kelly}}, \bibinfo {author}
  {\bibfnamefont {A.}~\bibnamefont {Megrant}}, \bibinfo {author} {\bibfnamefont
  {C.}~\bibnamefont {Neill}}, \bibinfo {author} {\bibfnamefont {P.~J.~J.}\
  \bibnamefont {O'Malley}}, \bibinfo {author} {\bibfnamefont {P.}~\bibnamefont
  {Roushan}}, \bibinfo {author} {\bibfnamefont {D.}~\bibnamefont {Sank}},
  \bibinfo {author} {\bibfnamefont {A.}~\bibnamefont {Vainsencher}}, \bibinfo
  {author} {\bibfnamefont {J.}~\bibnamefont {Wenner}}, \bibinfo {author}
  {\bibfnamefont {S.}~\bibnamefont {Chaudhuri}}, \bibinfo {author}
  {\bibfnamefont {J.}~\bibnamefont {Gao}},\ and\ \bibinfo {author}
  {\bibfnamefont {J.~M.}\ \bibnamefont {Martinis}},\ }\bibfield  {title}
  {\bibinfo {title} {Traveling wave parametric amplifier with {J}osephson
  junctions using minimal resonator phase matching},\ }\href
  {https://doi.org/10.1063/1.4922348} {\bibfield  {journal} {\bibinfo
  {journal} {Appl. Phys. Lett.}\ }\textbf {\bibinfo {volume} {106}},\ \bibinfo
  {pages} {242601} (\bibinfo {year} {2015})}\BibitemShut {NoStop}%
\bibitem [{\citenamefont {Perelshtein}\ \emph {et~al.}(2022)\citenamefont
  {Perelshtein}, \citenamefont {Petrovnin}, \citenamefont {Vesterinen},
  \citenamefont {Hamedani~Raja}, \citenamefont {Lilja}, \citenamefont {Will},
  \citenamefont {Savin}, \citenamefont {Simbierowicz}, \citenamefont
  {Jabdaraghi}, \citenamefont {Lehtinen}, \citenamefont {Gr\"onberg},
  \citenamefont {Hassel}, \citenamefont {Prunnila}, \citenamefont {Govenius},
  \citenamefont {Paraoanu},\ and\ \citenamefont {Hakonen}}]{Perelshtein2022}%
  \BibitemOpen
  \bibfield  {author} {\bibinfo {author} {\bibfnamefont {M.~R.}\ \bibnamefont
  {Perelshtein}}, \bibinfo {author} {\bibfnamefont {K.~V.}\ \bibnamefont
  {Petrovnin}}, \bibinfo {author} {\bibfnamefont {V.}~\bibnamefont
  {Vesterinen}}, \bibinfo {author} {\bibfnamefont {S.}~\bibnamefont
  {Hamedani~Raja}}, \bibinfo {author} {\bibfnamefont {I.}~\bibnamefont
  {Lilja}}, \bibinfo {author} {\bibfnamefont {M.}~\bibnamefont {Will}},
  \bibinfo {author} {\bibfnamefont {A.}~\bibnamefont {Savin}}, \bibinfo
  {author} {\bibfnamefont {S.}~\bibnamefont {Simbierowicz}}, \bibinfo {author}
  {\bibfnamefont {R.~N.}\ \bibnamefont {Jabdaraghi}}, \bibinfo {author}
  {\bibfnamefont {J.~S.}\ \bibnamefont {Lehtinen}}, \bibinfo {author}
  {\bibfnamefont {L.}~\bibnamefont {Gr\"onberg}}, \bibinfo {author}
  {\bibfnamefont {J.}~\bibnamefont {Hassel}}, \bibinfo {author} {\bibfnamefont
  {M.~P.}\ \bibnamefont {Prunnila}}, \bibinfo {author} {\bibfnamefont
  {J.}~\bibnamefont {Govenius}}, \bibinfo {author} {\bibfnamefont {G.~S.}\
  \bibnamefont {Paraoanu}},\ and\ \bibinfo {author} {\bibfnamefont {P.~J.}\
  \bibnamefont {Hakonen}},\ }\bibfield  {title} {\bibinfo {title} {Broadband
  continuous-variable entanglement generation using a {K}err-free {J}osephson
  metamaterial},\ }\href {https://doi.org/10.1103/PhysRevApplied.18.024063}
  {\bibfield  {journal} {\bibinfo  {journal} {Phys. Rev. Appl.}\ }\textbf
  {\bibinfo {volume} {18}},\ \bibinfo {pages} {024063} (\bibinfo {year}
  {2022})}\BibitemShut {NoStop}%
\bibitem [{\citenamefont {Planat}\ \emph {et~al.}(2020)\citenamefont {Planat},
  \citenamefont {Ranadive}, \citenamefont {Dassonneville}, \citenamefont
  {Puertas~Mart\'inez}, \citenamefont {L\'eger}, \citenamefont {Naud},
  \citenamefont {Buisson}, \citenamefont {Hasch-Guichard}, \citenamefont
  {Basko},\ and\ \citenamefont {Roch}}]{Planat2020}%
  \BibitemOpen
  \bibfield  {author} {\bibinfo {author} {\bibfnamefont {L.}~\bibnamefont
  {Planat}}, \bibinfo {author} {\bibfnamefont {A.}~\bibnamefont {Ranadive}},
  \bibinfo {author} {\bibfnamefont {R.}~\bibnamefont {Dassonneville}}, \bibinfo
  {author} {\bibfnamefont {J.}~\bibnamefont {Puertas~Mart\'inez}}, \bibinfo
  {author} {\bibfnamefont {S.}~\bibnamefont {L\'eger}}, \bibinfo {author}
  {\bibfnamefont {C.}~\bibnamefont {Naud}}, \bibinfo {author} {\bibfnamefont
  {O.}~\bibnamefont {Buisson}}, \bibinfo {author} {\bibfnamefont
  {W.}~\bibnamefont {Hasch-Guichard}}, \bibinfo {author} {\bibfnamefont
  {D.~M.}\ \bibnamefont {Basko}},\ and\ \bibinfo {author} {\bibfnamefont
  {N.}~\bibnamefont {Roch}},\ }\bibfield  {title} {\bibinfo {title}
  {Photonic-crystal {J}osephson traveling-wave parametric amplifier},\ }\href
  {https://doi.org/10.1103/PhysRevX.10.021021} {\bibfield  {journal} {\bibinfo
  {journal} {Phys. Rev. X}\ }\textbf {\bibinfo {volume} {10}},\ \bibinfo
  {pages} {021021} (\bibinfo {year} {2020})}\BibitemShut {NoStop}%
\bibitem [{\citenamefont {Chang}\ \emph {et~al.}(2025)\citenamefont {Chang},
  \citenamefont {Van~Loo}, \citenamefont {Hung}, \citenamefont {Zhou},
  \citenamefont {Gnandt}, \citenamefont {Tamate},\ and\ \citenamefont
  {Nakamura}}]{Chang2025}%
  \BibitemOpen
  \bibfield  {author} {\bibinfo {author} {\bibfnamefont {C.~W.~S.}\
  \bibnamefont {Chang}}, \bibinfo {author} {\bibfnamefont {A.~F.}\ \bibnamefont
  {Van~Loo}}, \bibinfo {author} {\bibfnamefont {C.-C.}\ \bibnamefont {Hung}},
  \bibinfo {author} {\bibfnamefont {Y.}~\bibnamefont {Zhou}}, \bibinfo {author}
  {\bibfnamefont {C.}~\bibnamefont {Gnandt}}, \bibinfo {author} {\bibfnamefont
  {S.}~\bibnamefont {Tamate}},\ and\ \bibinfo {author} {\bibfnamefont
  {Y.}~\bibnamefont {Nakamura}},\ }\bibfield  {title} {\bibinfo {title}
  {{J}osephson traveling-wave parametric amplifier based on a
  low-intrinsic-loss lumped-element coplanar waveguide},\ }\href
  {https://doi.org/10.1103/qhl6-cz2z} {\bibfield  {journal} {\bibinfo
  {journal} {Phys. Rev. Appl.}\ }\textbf {\bibinfo {volume} {24}},\ \bibinfo
  {pages} {044081} (\bibinfo {year} {2025})}\BibitemShut {NoStop}%
\bibitem [{\citenamefont {Fadavi~Roudsari}\ \emph {et~al.}(2023)\citenamefont
  {Fadavi~Roudsari}, \citenamefont {Shiri}, \citenamefont {Renberg~Nilsson},
  \citenamefont {Tancredi}, \citenamefont {Osman}, \citenamefont {Svensson},
  \citenamefont {Kudra}, \citenamefont {Rommel}, \citenamefont {Bylander},
  \citenamefont {Shumeiko},\ and\ \citenamefont
  {Delsing}}]{FadaviRoudsari2023}%
  \BibitemOpen
  \bibfield  {author} {\bibinfo {author} {\bibfnamefont {A.}~\bibnamefont
  {Fadavi~Roudsari}}, \bibinfo {author} {\bibfnamefont {D.}~\bibnamefont
  {Shiri}}, \bibinfo {author} {\bibfnamefont {H.}~\bibnamefont
  {Renberg~Nilsson}}, \bibinfo {author} {\bibfnamefont {G.}~\bibnamefont
  {Tancredi}}, \bibinfo {author} {\bibfnamefont {A.}~\bibnamefont {Osman}},
  \bibinfo {author} {\bibfnamefont {I.-M.}\ \bibnamefont {Svensson}}, \bibinfo
  {author} {\bibfnamefont {M.}~\bibnamefont {Kudra}}, \bibinfo {author}
  {\bibfnamefont {M.}~\bibnamefont {Rommel}}, \bibinfo {author} {\bibfnamefont
  {J.}~\bibnamefont {Bylander}}, \bibinfo {author} {\bibfnamefont
  {V.}~\bibnamefont {Shumeiko}},\ and\ \bibinfo {author} {\bibfnamefont
  {P.}~\bibnamefont {Delsing}},\ }\bibfield  {title} {\bibinfo {title}
  {Three-wave mixing traveling-wave parametric amplifier with periodic
  variation of the circuit parameters},\ }\href
  {https://doi.org/10.1063/5.0127690} {\bibfield  {journal} {\bibinfo
  {journal} {Appl. Phys. Lett.}\ }\textbf {\bibinfo {volume} {122}},\ \bibinfo
  {pages} {052601} (\bibinfo {year} {2023})}\BibitemShut {NoStop}%
\bibitem [{\citenamefont {Ranadive}\ \emph {et~al.}(2022)\citenamefont
  {Ranadive}, \citenamefont {Esposito}, \citenamefont {Planat}, \citenamefont
  {Bonet}, \citenamefont {Naud}, \citenamefont {Buisson}, \citenamefont
  {Guichard},\ and\ \citenamefont {Roch}}]{Ranadive2022}%
  \BibitemOpen
  \bibfield  {author} {\bibinfo {author} {\bibfnamefont {A.}~\bibnamefont
  {Ranadive}}, \bibinfo {author} {\bibfnamefont {M.}~\bibnamefont {Esposito}},
  \bibinfo {author} {\bibfnamefont {L.}~\bibnamefont {Planat}}, \bibinfo
  {author} {\bibfnamefont {E.}~\bibnamefont {Bonet}}, \bibinfo {author}
  {\bibfnamefont {C.}~\bibnamefont {Naud}}, \bibinfo {author} {\bibfnamefont
  {O.}~\bibnamefont {Buisson}}, \bibinfo {author} {\bibfnamefont
  {W.}~\bibnamefont {Guichard}},\ and\ \bibinfo {author} {\bibfnamefont
  {N.}~\bibnamefont {Roch}},\ }\bibfield  {title} {\bibinfo {title} {{K}err
  reversal in {J}osephson meta-material and traveling wave parametric
  amplification},\ }\href {https://doi.org/10.1038/s41467-022-29375-5}
  {\bibfield  {journal} {\bibinfo  {journal} {Nat. Commun.}\ }\textbf {\bibinfo
  {volume} {13}},\ \bibinfo {pages} {1737} (\bibinfo {year}
  {2022})}\BibitemShut {NoStop}%
\bibitem [{\citenamefont {Bell}(2026)}]{Bell2026}%
  \BibitemOpen
  \bibfield  {author} {\bibinfo {author} {\bibfnamefont {M.~T.}\ \bibnamefont
  {Bell}},\ }\bibfield  {title} {\bibinfo {title} {{J}osephson traveling-wave
  parametric amplifier with inverse {K}err phase matching},\ }\href
  {https://doi.org/10.1103/frfk-ty7x} {\bibfield  {journal} {\bibinfo
  {journal} {Phys. Rev. Appl.}\ }\textbf {\bibinfo {volume} {25}},\ \bibinfo
  {pages} {014075} (\bibinfo {year} {2026})}\BibitemShut {NoStop}%
\bibitem [{\citenamefont {Kow}\ \emph {et~al.}(2025)\citenamefont {Kow},
  \citenamefont {Podolskiy},\ and\ \citenamefont {Kamal}}]{Kow2025}%
  \BibitemOpen
  \bibfield  {author} {\bibinfo {author} {\bibfnamefont {C.}~\bibnamefont
  {Kow}}, \bibinfo {author} {\bibfnamefont {V.}~\bibnamefont {Podolskiy}},\
  and\ \bibinfo {author} {\bibfnamefont {A.}~\bibnamefont {Kamal}},\ }\bibfield
   {title} {\bibinfo {title} {Self-phase-matched broadband amplification with a
  left-handed {J}osephson transmission line},\ }\href
  {https://doi.org/10.1103/PhysRevApplied.24.024026} {\bibfield  {journal}
  {\bibinfo  {journal} {Phys. Rev. Appl.}\ }\textbf {\bibinfo {volume} {24}},\
  \bibinfo {pages} {024026} (\bibinfo {year} {2025})}\BibitemShut {NoStop}%
\bibitem [{\citenamefont {Zorin}(2016)}]{Zorin2016}%
  \BibitemOpen
  \bibfield  {author} {\bibinfo {author} {\bibfnamefont {A.~B.}\ \bibnamefont
  {Zorin}},\ }\bibfield  {title} {\bibinfo {title} {Josephson traveling-wave
  parametric amplifier with three-wave mixing},\ }\href
  {https://doi.org/10.1103/physrevapplied.6.034006} {\bibfield  {journal}
  {\bibinfo  {journal} {Phys. Rev. Appl.}\ }\textbf {\bibinfo {volume} {6}},\
  \bibinfo {pages} {034006} (\bibinfo {year} {2016})}\BibitemShut {NoStop}%
\bibitem [{\citenamefont {Gaydamachenko}\ \emph {et~al.}(2025)\citenamefont
  {Gaydamachenko}, \citenamefont {Kissling},\ and\ \citenamefont
  {Gr\"unhaupt}}]{Gaydamachenko2025}%
  \BibitemOpen
  \bibfield  {author} {\bibinfo {author} {\bibfnamefont {V.}~\bibnamefont
  {Gaydamachenko}}, \bibinfo {author} {\bibfnamefont {C.}~\bibnamefont
  {Kissling}},\ and\ \bibinfo {author} {\bibfnamefont {L.}~\bibnamefont
  {Gr\"unhaupt}},\ }\bibfield  {title} {\bibinfo {title} {{rf}-{SQUID}-based
  traveling-wave parametric amplifier with input saturation power of $-84$
  {dBm} across more than one octave in bandwidth},\ }\href
  {https://doi.org/10.1103/1qk4-fzkq} {\bibfield  {journal} {\bibinfo
  {journal} {Phys. Rev. Appl.}\ }\textbf {\bibinfo {volume} {23}},\ \bibinfo
  {pages} {064053} (\bibinfo {year} {2025})}\BibitemShut {NoStop}%
\bibitem [{\citenamefont {Livreri}\ \emph {et~al.}(2024)\citenamefont
  {Livreri}, \citenamefont {Galvano}, \citenamefont {Fasolo}, \citenamefont
  {Oberto},\ and\ \citenamefont {Enrico}}]{Livreri2024}%
  \BibitemOpen
  \bibfield  {author} {\bibinfo {author} {\bibfnamefont {P.}~\bibnamefont
  {Livreri}}, \bibinfo {author} {\bibfnamefont {B.}~\bibnamefont {Galvano}},
  \bibinfo {author} {\bibfnamefont {L.}~\bibnamefont {Fasolo}}, \bibinfo
  {author} {\bibfnamefont {L.}~\bibnamefont {Oberto}},\ and\ \bibinfo {author}
  {\bibfnamefont {E.}~\bibnamefont {Enrico}},\ }\bibfield  {title} {\bibinfo
  {title} {{J}osephson traveling wave parametric amplifier as quantum source of
  entangled photons for microwave quantum radar applications},\ }\href
  {https://doi.org/10.2528/pier24041705} {\bibfield  {journal} {\bibinfo
  {journal} {Prog. Electromagn. Res.}\ }\textbf {\bibinfo {volume} {179}},\
  \bibinfo {pages} {113} (\bibinfo {year} {2024})}\BibitemShut {NoStop}%
\bibitem [{\citenamefont {Di~Vora}\ \emph {et~al.}(2023)\citenamefont
  {Di~Vora}, \citenamefont {Lombardi}, \citenamefont {Ortolan}, \citenamefont
  {Pengo}, \citenamefont {Ruoso}, \citenamefont {Braggio}, \citenamefont
  {Carugno}, \citenamefont {Taffarello}, \citenamefont {Cappelli},
  \citenamefont {Crescini}, \citenamefont {Esposito}, \citenamefont {Planat},
  \citenamefont {Ranadive}, \citenamefont {Roch}, \citenamefont {Alesini},
  \citenamefont {Babusci}, \citenamefont {D'Elia}, \citenamefont
  {Di~Gioacchino}, \citenamefont {Gatti}, \citenamefont {Ligi}, \citenamefont
  {Maccarrone}, \citenamefont {Rettaroli}, \citenamefont {Tocci}, \citenamefont
  {D'Agostino}, \citenamefont {Gambardella}, \citenamefont {Iannone},
  \citenamefont {Falferi} \emph {et~al.}}]{DiVora2023}%
  \BibitemOpen
  \bibfield  {author} {\bibinfo {author} {\bibfnamefont {R.}~\bibnamefont
  {Di~Vora}}, \bibinfo {author} {\bibfnamefont {A.}~\bibnamefont {Lombardi}},
  \bibinfo {author} {\bibfnamefont {A.}~\bibnamefont {Ortolan}}, \bibinfo
  {author} {\bibfnamefont {R.}~\bibnamefont {Pengo}}, \bibinfo {author}
  {\bibfnamefont {G.}~\bibnamefont {Ruoso}}, \bibinfo {author} {\bibfnamefont
  {C.}~\bibnamefont {Braggio}}, \bibinfo {author} {\bibfnamefont
  {G.}~\bibnamefont {Carugno}}, \bibinfo {author} {\bibfnamefont
  {L.}~\bibnamefont {Taffarello}}, \bibinfo {author} {\bibfnamefont
  {G.}~\bibnamefont {Cappelli}}, \bibinfo {author} {\bibfnamefont
  {N.}~\bibnamefont {Crescini}}, \bibinfo {author} {\bibfnamefont
  {M.}~\bibnamefont {Esposito}}, \bibinfo {author} {\bibfnamefont
  {L.}~\bibnamefont {Planat}}, \bibinfo {author} {\bibfnamefont
  {A.}~\bibnamefont {Ranadive}}, \bibinfo {author} {\bibfnamefont
  {N.}~\bibnamefont {Roch}}, \bibinfo {author} {\bibfnamefont {D.}~\bibnamefont
  {Alesini}}, \bibinfo {author} {\bibfnamefont {D.}~\bibnamefont {Babusci}},
  \bibinfo {author} {\bibfnamefont {A.}~\bibnamefont {D'Elia}}, \bibinfo
  {author} {\bibfnamefont {D.}~\bibnamefont {Di~Gioacchino}}, \bibinfo {author}
  {\bibfnamefont {C.}~\bibnamefont {Gatti}}, \bibinfo {author} {\bibfnamefont
  {C.}~\bibnamefont {Ligi}}, \bibinfo {author} {\bibfnamefont {G.}~\bibnamefont
  {Maccarrone}}, \bibinfo {author} {\bibfnamefont {A.}~\bibnamefont
  {Rettaroli}}, \bibinfo {author} {\bibfnamefont {S.}~\bibnamefont {Tocci}},
  \bibinfo {author} {\bibfnamefont {D.}~\bibnamefont {D'Agostino}}, \bibinfo
  {author} {\bibfnamefont {U.}~\bibnamefont {Gambardella}}, \bibinfo {author}
  {\bibfnamefont {G.}~\bibnamefont {Iannone}}, \bibinfo {author} {\bibfnamefont
  {P.}~\bibnamefont {Falferi}}, \emph {et~al.},\ }\bibfield  {title} {\bibinfo
  {title} {Search for galactic axions with a traveling wave parametric
  amplifier},\ }\href {https://doi.org/10.1103/PhysRevD.108.062005} {\bibfield
  {journal} {\bibinfo  {journal} {Phys. Rev. D}\ }\textbf {\bibinfo {volume}
  {108}},\ \bibinfo {pages} {062005} (\bibinfo {year} {2023})}\BibitemShut
  {NoStop}%
\bibitem [{\citenamefont {Kow}\ and\ \citenamefont {Bell}(2026)}]{Kow2026}%
  \BibitemOpen
  \bibfield  {author} {\bibinfo {author} {\bibfnamefont {C.~S.}\ \bibnamefont
  {Kow}}\ and\ \bibinfo {author} {\bibfnamefont {M.~T.}\ \bibnamefont {Bell}},\
  }\bibfield  {title} {\bibinfo {title} {Traveling-wave parametric amplifier
  with passive reverse isolation},\ }\href {https://doi.org/10.1103/zkyt-r8vd}
  {\bibfield  {journal} {\bibinfo  {journal} {Phys. Rev. X}\ }\textbf {\bibinfo
  {volume} {16}},\ \bibinfo {pages} {021003} (\bibinfo {year}
  {2026})}\BibitemShut {NoStop}%
\bibitem [{\citenamefont {Zapata}\ \emph {et~al.}(1996)\citenamefont {Zapata},
  \citenamefont {Bartussek}, \citenamefont {Sols},\ and\ \citenamefont
  {H\"anggi}}]{Zapata1996}%
  \BibitemOpen
  \bibfield  {author} {\bibinfo {author} {\bibfnamefont {I.}~\bibnamefont
  {Zapata}}, \bibinfo {author} {\bibfnamefont {R.}~\bibnamefont {Bartussek}},
  \bibinfo {author} {\bibfnamefont {F.}~\bibnamefont {Sols}},\ and\ \bibinfo
  {author} {\bibfnamefont {P.}~\bibnamefont {H\"anggi}},\ }\bibfield  {title}
  {\bibinfo {title} {Voltage rectification by a {SQUID} ratchet},\ }\href
  {https://doi.org/10.1103/PhysRevLett.77.2292} {\bibfield  {journal} {\bibinfo
   {journal} {Phys. Rev. Lett.}\ }\textbf {\bibinfo {volume} {77}},\ \bibinfo
  {pages} {2292} (\bibinfo {year} {1996})}\BibitemShut {NoStop}%
\bibitem [{\citenamefont {Haider}(2024)}]{Haider2024}%
  \BibitemOpen
  \bibfield  {author} {\bibinfo {author} {\bibfnamefont {M.}~\bibnamefont
  {Haider}},\ }\bibfield  {title} {\bibinfo {title} {A black-box quantum model
  for superconducting traveling-wave parametric amplifiers},\ }\href
  {https://doi.org/10.1109/TMTT.2023.3345641} {\bibfield  {journal} {\bibinfo
  {journal} {IEEE Trans. Microwave Theory Tech.}\ }\textbf {\bibinfo {volume}
  {72}},\ \bibinfo {pages} {2143} (\bibinfo {year} {2024})}\BibitemShut
  {NoStop}%
\bibitem [{\citenamefont {Sivak}\ \emph {et~al.}(2019)\citenamefont {Sivak},
  \citenamefont {Frattini}, \citenamefont {Joshi}, \citenamefont
  {Lingenfelter}, \citenamefont {Shankar},\ and\ \citenamefont
  {Devoret}}]{Sivak2019}%
  \BibitemOpen
  \bibfield  {author} {\bibinfo {author} {\bibfnamefont {V.}~\bibnamefont
  {Sivak}}, \bibinfo {author} {\bibfnamefont {N.}~\bibnamefont {Frattini}},
  \bibinfo {author} {\bibfnamefont {V.}~\bibnamefont {Joshi}}, \bibinfo
  {author} {\bibfnamefont {A.}~\bibnamefont {Lingenfelter}}, \bibinfo {author}
  {\bibfnamefont {S.}~\bibnamefont {Shankar}},\ and\ \bibinfo {author}
  {\bibfnamefont {M.}~\bibnamefont {Devoret}},\ }\bibfield  {title} {\bibinfo
  {title} {Kerr-free three-wave mixing in superconducting quantum circuits},\
  }\href {https://doi.org/10.1103/physrevapplied.11.054060} {\bibfield
  {journal} {\bibinfo  {journal} {Phys. Rev. Appl.}\ }\textbf {\bibinfo
  {volume} {11}},\ \bibinfo {pages} {054060} (\bibinfo {year}
  {2019})}\BibitemShut {NoStop}%
\bibitem [{\citenamefont {Roy}\ \emph {et~al.}(2025)\citenamefont {Roy},
  \citenamefont {Romeiro}, \citenamefont {Koch}, \citenamefont {Tsitsilin},
  \citenamefont {Schirk}, \citenamefont {Glaser}, \citenamefont {Bruckmoser},
  \citenamefont {Singh}, \citenamefont {Haslbeck}, \citenamefont {Huber},
  \citenamefont {Krylov}, \citenamefont {Marx}, \citenamefont {Pfeiffer},
  \citenamefont {Schneider}, \citenamefont {Schweizer}, \citenamefont
  {Wallner}, \citenamefont {Bunch}, \citenamefont {Richard}, \citenamefont
  {S{\"o}dergren}, \citenamefont {Liegener}, \citenamefont {Werninghaus},\ and\
  \citenamefont {Filipp}}]{Roy2025}%
  \BibitemOpen
  \bibfield  {author} {\bibinfo {author} {\bibfnamefont {F.~A.}\ \bibnamefont
  {Roy}}, \bibinfo {author} {\bibfnamefont {J.~H.}\ \bibnamefont {Romeiro}},
  \bibinfo {author} {\bibfnamefont {L.}~\bibnamefont {Koch}}, \bibinfo {author}
  {\bibfnamefont {I.}~\bibnamefont {Tsitsilin}}, \bibinfo {author}
  {\bibfnamefont {J.}~\bibnamefont {Schirk}}, \bibinfo {author} {\bibfnamefont
  {N.~J.}\ \bibnamefont {Glaser}}, \bibinfo {author} {\bibfnamefont
  {N.}~\bibnamefont {Bruckmoser}}, \bibinfo {author} {\bibfnamefont
  {M.}~\bibnamefont {Singh}}, \bibinfo {author} {\bibfnamefont {F.~X.}\
  \bibnamefont {Haslbeck}}, \bibinfo {author} {\bibfnamefont {G.~B.~P.}\
  \bibnamefont {Huber}}, \bibinfo {author} {\bibfnamefont {G.}~\bibnamefont
  {Krylov}}, \bibinfo {author} {\bibfnamefont {A.}~\bibnamefont {Marx}},
  \bibinfo {author} {\bibfnamefont {F.}~\bibnamefont {Pfeiffer}}, \bibinfo
  {author} {\bibfnamefont {C.~M.~F.}\ \bibnamefont {Schneider}}, \bibinfo
  {author} {\bibfnamefont {C.}~\bibnamefont {Schweizer}}, \bibinfo {author}
  {\bibfnamefont {F.}~\bibnamefont {Wallner}}, \bibinfo {author} {\bibfnamefont
  {D.}~\bibnamefont {Bunch}}, \bibinfo {author} {\bibfnamefont
  {L.}~\bibnamefont {Richard}}, \bibinfo {author} {\bibfnamefont
  {L.}~\bibnamefont {S{\"o}dergren}}, \bibinfo {author} {\bibfnamefont
  {K.}~\bibnamefont {Liegener}}, \bibinfo {author} {\bibfnamefont
  {M.}~\bibnamefont {Werninghaus}},\ and\ \bibinfo {author} {\bibfnamefont
  {S.}~\bibnamefont {Filipp}},\ }\bibfield  {title} {\bibinfo {title}
  {Parity-dependent state transfer for direct entanglement generation},\ }\href
  {https://doi.org/10.1038/s41467-025-57818-2} {\bibfield  {journal} {\bibinfo
  {journal} {Nat. Commun.}\ }\textbf {\bibinfo {volume} {16}},\ \bibinfo
  {pages} {2660} (\bibinfo {year} {2025})}\BibitemShut {NoStop}%
\bibitem [{\citenamefont {Honasoge}\ \emph {et~al.}(2025)\citenamefont
  {Honasoge}, \citenamefont {Handschuh}, \citenamefont {Yam}, \citenamefont
  {Gandorfer}, \citenamefont {Bazulin}, \citenamefont {Bruckmoser},
  \citenamefont {Koch}, \citenamefont {Marx}, \citenamefont {Gross},\ and\
  \citenamefont {Fedorov}}]{Honasoge2025}%
  \BibitemOpen
  \bibfield  {author} {\bibinfo {author} {\bibfnamefont {K.~E.}\ \bibnamefont
  {Honasoge}}, \bibinfo {author} {\bibfnamefont {M.}~\bibnamefont {Handschuh}},
  \bibinfo {author} {\bibfnamefont {W.~K.}\ \bibnamefont {Yam}}, \bibinfo
  {author} {\bibfnamefont {S.}~\bibnamefont {Gandorfer}}, \bibinfo {author}
  {\bibfnamefont {D.}~\bibnamefont {Bazulin}}, \bibinfo {author} {\bibfnamefont
  {N.}~\bibnamefont {Bruckmoser}}, \bibinfo {author} {\bibfnamefont
  {L.}~\bibnamefont {Koch}}, \bibinfo {author} {\bibfnamefont {A.}~\bibnamefont
  {Marx}}, \bibinfo {author} {\bibfnamefont {R.}~\bibnamefont {Gross}},\ and\
  \bibinfo {author} {\bibfnamefont {K.~G.}\ \bibnamefont {Fedorov}},\
  }\bibfield  {title} {\bibinfo {title} {Fabrication of low-loss {J}osephson
  parametric devices},\ }\href {https://doi.org/10.1103/PhysRevB.111.214508}
  {\bibfield  {journal} {\bibinfo  {journal} {Phys. Rev. B}\ }\textbf {\bibinfo
  {volume} {111}},\ \bibinfo {pages} {214508} (\bibinfo {year}
  {2025})}\BibitemShut {NoStop}%
\bibitem [{\citenamefont {Lu}\ \emph {et~al.}(2022)\citenamefont {Lu},
  \citenamefont {Xu}, \citenamefont {Zuo}, \citenamefont {Pan}, \citenamefont
  {Wei}, \citenamefont {Jiang}, \citenamefont {Li}, \citenamefont {Zhang},
  \citenamefont {Guo}, \citenamefont {Wang}, \citenamefont {Cao}, \citenamefont
  {Wang}, \citenamefont {Xu}, \citenamefont {Sun},\ and\ \citenamefont
  {Wu}}]{Lu2022}%
  \BibitemOpen
  \bibfield  {author} {\bibinfo {author} {\bibfnamefont {Y.}~\bibnamefont
  {Lu}}, \bibinfo {author} {\bibfnamefont {W.}~\bibnamefont {Xu}}, \bibinfo
  {author} {\bibfnamefont {Q.}~\bibnamefont {Zuo}}, \bibinfo {author}
  {\bibfnamefont {J.}~\bibnamefont {Pan}}, \bibinfo {author} {\bibfnamefont
  {X.}~\bibnamefont {Wei}}, \bibinfo {author} {\bibfnamefont {J.}~\bibnamefont
  {Jiang}}, \bibinfo {author} {\bibfnamefont {Z.}~\bibnamefont {Li}}, \bibinfo
  {author} {\bibfnamefont {K.}~\bibnamefont {Zhang}}, \bibinfo {author}
  {\bibfnamefont {T.}~\bibnamefont {Guo}}, \bibinfo {author} {\bibfnamefont
  {S.}~\bibnamefont {Wang}}, \bibinfo {author} {\bibfnamefont {C.}~\bibnamefont
  {Cao}}, \bibinfo {author} {\bibfnamefont {H.}~\bibnamefont {Wang}}, \bibinfo
  {author} {\bibfnamefont {W.}~\bibnamefont {Xu}}, \bibinfo {author}
  {\bibfnamefont {G.}~\bibnamefont {Sun}},\ and\ \bibinfo {author}
  {\bibfnamefont {P.}~\bibnamefont {Wu}},\ }\bibfield  {title} {\bibinfo
  {title} {Broadband {J}osephson parametric amplifier using lumped-element
  transmission line impedance matching architecture},\ }\href
  {https://doi.org/10.1063/5.0080314} {\bibfield  {journal} {\bibinfo
  {journal} {Appl. Phys. Lett.}\ }\textbf {\bibinfo {volume} {120}},\ \bibinfo
  {pages} {082601} (\bibinfo {year} {2022})}\BibitemShut {NoStop}%
\bibitem [{\citenamefont {Chaudhuri}\ \emph {et~al.}(2017)\citenamefont
  {Chaudhuri}, \citenamefont {Li}, \citenamefont {Irwin}, \citenamefont
  {Bockstiegel}, \citenamefont {Hubmayr}, \citenamefont {Ullom}, \citenamefont
  {Vissers},\ and\ \citenamefont {Gao}}]{Chaudhuri2017}%
  \BibitemOpen
  \bibfield  {author} {\bibinfo {author} {\bibfnamefont {S.}~\bibnamefont
  {Chaudhuri}}, \bibinfo {author} {\bibfnamefont {D.}~\bibnamefont {Li}},
  \bibinfo {author} {\bibfnamefont {K.~D.}\ \bibnamefont {Irwin}}, \bibinfo
  {author} {\bibfnamefont {C.}~\bibnamefont {Bockstiegel}}, \bibinfo {author}
  {\bibfnamefont {J.}~\bibnamefont {Hubmayr}}, \bibinfo {author} {\bibfnamefont
  {J.~N.}\ \bibnamefont {Ullom}}, \bibinfo {author} {\bibfnamefont {M.~R.}\
  \bibnamefont {Vissers}},\ and\ \bibinfo {author} {\bibfnamefont
  {J.}~\bibnamefont {Gao}},\ }\bibfield  {title} {\bibinfo {title} {Broadband
  parametric amplifiers based on nonlinear kinetic inductance artificial
  transmission lines},\ }\href {https://doi.org/10.1063/1.4980102} {\bibfield
  {journal} {\bibinfo  {journal} {Appl. Phys. Lett.}\ }\textbf {\bibinfo
  {volume} {110}},\ \bibinfo {pages} {152601} (\bibinfo {year}
  {2017})}\BibitemShut {NoStop}%
\bibitem [{\citenamefont {Malnou}\ \emph {et~al.}(2021)\citenamefont {Malnou},
  \citenamefont {Vissers}, \citenamefont {Wheeler}, \citenamefont {Aumentado},
  \citenamefont {Hubmayr}, \citenamefont {Ullom},\ and\ \citenamefont
  {Gao}}]{Malnou2021}%
  \BibitemOpen
  \bibfield  {author} {\bibinfo {author} {\bibfnamefont {M.}~\bibnamefont
  {Malnou}}, \bibinfo {author} {\bibfnamefont {M.~R.}\ \bibnamefont {Vissers}},
  \bibinfo {author} {\bibfnamefont {J.~D.}\ \bibnamefont {Wheeler}}, \bibinfo
  {author} {\bibfnamefont {J.}~\bibnamefont {Aumentado}}, \bibinfo {author}
  {\bibfnamefont {J.}~\bibnamefont {Hubmayr}}, \bibinfo {author} {\bibfnamefont
  {J.~N.}\ \bibnamefont {Ullom}},\ and\ \bibinfo {author} {\bibfnamefont
  {J.}~\bibnamefont {Gao}},\ }\bibfield  {title} {\bibinfo {title} {Three-wave
  mixing kinetic inductance traveling-wave amplifier with near-quantum-limited
  noise performance},\ }\href {https://doi.org/10.1103/PRXQuantum.2.010302}
  {\bibfield  {journal} {\bibinfo  {journal} {PRX Quantum}\ }\textbf {\bibinfo
  {volume} {2}},\ \bibinfo {pages} {010302} (\bibinfo {year}
  {2021})}\BibitemShut {NoStop}%
\bibitem [{\citenamefont {Wang}\ \emph {et~al.}(2025)\citenamefont {Wang},
  \citenamefont {Lu}, \citenamefont {Zhan}, \citenamefont {Ma}, \citenamefont
  {Wu}, \citenamefont {Sun}, \citenamefont {Deng}, \citenamefont {Bai},
  \citenamefont {Bao}, \citenamefont {Chang}, \citenamefont {Gao},
  \citenamefont {Gao}, \citenamefont {Gong}, \citenamefont {Hu}, \citenamefont
  {Hu}, \citenamefont {Ji}, \citenamefont {Ma}, \citenamefont {Mao},
  \citenamefont {Song}, \citenamefont {Tang}, \citenamefont {Wang},
  \citenamefont {Wang}, \citenamefont {Wang}, \citenamefont {Xia},
  \citenamefont {Xu}, \citenamefont {Zhan}, \citenamefont {Zhang},
  \citenamefont {Zhou}, \citenamefont {Zhu}, \citenamefont {Zhu}, \citenamefont
  {Zhu}, \citenamefont {Zhu}, \citenamefont {Shi}, \citenamefont {Zhao},\ and\
  \citenamefont {Deng}}]{Wang2025}%
  \BibitemOpen
  \bibfield  {author} {\bibinfo {author} {\bibfnamefont {F.}~\bibnamefont
  {Wang}}, \bibinfo {author} {\bibfnamefont {K.}~\bibnamefont {Lu}}, \bibinfo
  {author} {\bibfnamefont {H.}~\bibnamefont {Zhan}}, \bibinfo {author}
  {\bibfnamefont {L.}~\bibnamefont {Ma}}, \bibinfo {author} {\bibfnamefont
  {F.}~\bibnamefont {Wu}}, \bibinfo {author} {\bibfnamefont {H.}~\bibnamefont
  {Sun}}, \bibinfo {author} {\bibfnamefont {H.}~\bibnamefont {Deng}}, \bibinfo
  {author} {\bibfnamefont {Y.}~\bibnamefont {Bai}}, \bibinfo {author}
  {\bibfnamefont {F.}~\bibnamefont {Bao}}, \bibinfo {author} {\bibfnamefont
  {X.}~\bibnamefont {Chang}}, \bibinfo {author} {\bibfnamefont
  {R.}~\bibnamefont {Gao}}, \bibinfo {author} {\bibfnamefont {X.}~\bibnamefont
  {Gao}}, \bibinfo {author} {\bibfnamefont {G.}~\bibnamefont {Gong}}, \bibinfo
  {author} {\bibfnamefont {L.}~\bibnamefont {Hu}}, \bibinfo {author}
  {\bibfnamefont {R.}~\bibnamefont {Hu}}, \bibinfo {author} {\bibfnamefont
  {H.}~\bibnamefont {Ji}}, \bibinfo {author} {\bibfnamefont {X.}~\bibnamefont
  {Ma}}, \bibinfo {author} {\bibfnamefont {L.}~\bibnamefont {Mao}}, \bibinfo
  {author} {\bibfnamefont {Z.}~\bibnamefont {Song}}, \bibinfo {author}
  {\bibfnamefont {C.}~\bibnamefont {Tang}}, \bibinfo {author} {\bibfnamefont
  {H.}~\bibnamefont {Wang}}, \bibinfo {author} {\bibfnamefont {T.}~\bibnamefont
  {Wang}}, \bibinfo {author} {\bibfnamefont {Z.}~\bibnamefont {Wang}}, \bibinfo
  {author} {\bibfnamefont {T.}~\bibnamefont {Xia}}, \bibinfo {author}
  {\bibfnamefont {H.}~\bibnamefont {Xu}}, \bibinfo {author} {\bibfnamefont
  {Z.}~\bibnamefont {Zhan}}, \bibinfo {author} {\bibfnamefont {G.}~\bibnamefont
  {Zhang}}, \bibinfo {author} {\bibfnamefont {T.}~\bibnamefont {Zhou}},
  \bibinfo {author} {\bibfnamefont {M.}~\bibnamefont {Zhu}}, \bibinfo {author}
  {\bibfnamefont {Q.}~\bibnamefont {Zhu}}, \bibinfo {author} {\bibfnamefont
  {S.}~\bibnamefont {Zhu}}, \bibinfo {author} {\bibfnamefont {X.}~\bibnamefont
  {Zhu}}, \bibinfo {author} {\bibfnamefont {Y.}~\bibnamefont {Shi}}, \bibinfo
  {author} {\bibfnamefont {H.-H.}\ \bibnamefont {Zhao}},\ and\ \bibinfo
  {author} {\bibfnamefont {C.}~\bibnamefont {Deng}},\ }\bibfield  {title}
  {\bibinfo {title} {High-coherence fluxonium qubits manufactured with a
  wafer-scale-uniformity process},\ }\href
  {https://doi.org/10.1103/PhysRevApplied.23.044064} {\bibfield  {journal}
  {\bibinfo  {journal} {Phys. Rev. Appl.}\ }\textbf {\bibinfo {volume} {23}},\
  \bibinfo {pages} {044064} (\bibinfo {year} {2025})}\BibitemShut {NoStop}%
\bibitem [{\citenamefont {Mahboob}\ \emph {et~al.}(2026)\citenamefont
  {Mahboob}, \citenamefont {Sasaki},\ and\ \citenamefont
  {Takenaka}}]{Mahboob2026}%
  \BibitemOpen
  \bibfield  {author} {\bibinfo {author} {\bibfnamefont {I.}~\bibnamefont
  {Mahboob}}, \bibinfo {author} {\bibfnamefont {S.}~\bibnamefont {Sasaki}},\
  and\ \bibinfo {author} {\bibfnamefont {T.}~\bibnamefont {Takenaka}},\
  }\href@noop {} {\bibinfo {title} {A simple method to fabricate {J}osephson
  junctions}} (\bibinfo {year} {2026}),\ \Eprint
  {https://arxiv.org/abs/2604.25195} {arXiv:2604.25195 [quant-ph]} \BibitemShut
  {NoStop}%
\bibitem [{\citenamefont {Dunsworth}\ \emph {et~al.}(2017)\citenamefont
  {Dunsworth}, \citenamefont {Megrant}, \citenamefont {Quintana}, \citenamefont
  {Chen}, \citenamefont {Barends}, \citenamefont {Burkett}, \citenamefont
  {Foxen}, \citenamefont {Chen}, \citenamefont {Chiaro}, \citenamefont
  {Fowler}, \citenamefont {Graff}, \citenamefont {Jeffrey}, \citenamefont
  {Kelly}, \citenamefont {Lucero}, \citenamefont {Mutus}, \citenamefont
  {Neeley}, \citenamefont {Neill}, \citenamefont {Roushan}, \citenamefont
  {Sank}, \citenamefont {Vainsencher}, \citenamefont {Wenner}, \citenamefont
  {White},\ and\ \citenamefont {Martinis}}]{Dunsworth2017}%
  \BibitemOpen
  \bibfield  {author} {\bibinfo {author} {\bibfnamefont {A.}~\bibnamefont
  {Dunsworth}}, \bibinfo {author} {\bibfnamefont {A.}~\bibnamefont {Megrant}},
  \bibinfo {author} {\bibfnamefont {C.}~\bibnamefont {Quintana}}, \bibinfo
  {author} {\bibfnamefont {Z.}~\bibnamefont {Chen}}, \bibinfo {author}
  {\bibfnamefont {R.}~\bibnamefont {Barends}}, \bibinfo {author} {\bibfnamefont
  {B.}~\bibnamefont {Burkett}}, \bibinfo {author} {\bibfnamefont
  {B.}~\bibnamefont {Foxen}}, \bibinfo {author} {\bibfnamefont
  {Y.}~\bibnamefont {Chen}}, \bibinfo {author} {\bibfnamefont {B.}~\bibnamefont
  {Chiaro}}, \bibinfo {author} {\bibfnamefont {A.}~\bibnamefont {Fowler}},
  \bibinfo {author} {\bibfnamefont {R.}~\bibnamefont {Graff}}, \bibinfo
  {author} {\bibfnamefont {E.}~\bibnamefont {Jeffrey}}, \bibinfo {author}
  {\bibfnamefont {J.}~\bibnamefont {Kelly}}, \bibinfo {author} {\bibfnamefont
  {E.}~\bibnamefont {Lucero}}, \bibinfo {author} {\bibfnamefont {J.~Y.}\
  \bibnamefont {Mutus}}, \bibinfo {author} {\bibfnamefont {M.}~\bibnamefont
  {Neeley}}, \bibinfo {author} {\bibfnamefont {C.}~\bibnamefont {Neill}},
  \bibinfo {author} {\bibfnamefont {P.}~\bibnamefont {Roushan}}, \bibinfo
  {author} {\bibfnamefont {D.}~\bibnamefont {Sank}}, \bibinfo {author}
  {\bibfnamefont {A.}~\bibnamefont {Vainsencher}}, \bibinfo {author}
  {\bibfnamefont {J.}~\bibnamefont {Wenner}}, \bibinfo {author} {\bibfnamefont
  {T.~C.}\ \bibnamefont {White}},\ and\ \bibinfo {author} {\bibfnamefont
  {J.~M.}\ \bibnamefont {Martinis}},\ }\bibfield  {title} {\bibinfo {title}
  {Characterization and reduction of capacitive loss induced by sub-micron
  {J}osephson junction fabrication in superconducting qubits},\ }\href
  {https://doi.org/10.1063/1.4993577} {\bibfield  {journal} {\bibinfo
  {journal} {Appl. Phys. Lett.}\ }\textbf {\bibinfo {volume} {111}},\ \bibinfo
  {pages} {022601} (\bibinfo {year} {2017})}\BibitemShut {NoStop}%
\bibitem [{\citenamefont {Renberg~Nilsson}\ \emph {et~al.}(2023)\citenamefont
  {Renberg~Nilsson}, \citenamefont {Fadavi~Roudsari}, \citenamefont {Shiri},
  \citenamefont {Delsing},\ and\ \citenamefont
  {Shumeiko}}]{RenbergNilsson2023}%
  \BibitemOpen
  \bibfield  {author} {\bibinfo {author} {\bibfnamefont {H.}~\bibnamefont
  {Renberg~Nilsson}}, \bibinfo {author} {\bibfnamefont {A.}~\bibnamefont
  {Fadavi~Roudsari}}, \bibinfo {author} {\bibfnamefont {D.}~\bibnamefont
  {Shiri}}, \bibinfo {author} {\bibfnamefont {P.}~\bibnamefont {Delsing}},\
  and\ \bibinfo {author} {\bibfnamefont {V.}~\bibnamefont {Shumeiko}},\
  }\bibfield  {title} {\bibinfo {title} {High-gain traveling-wave parametric
  amplifier based on three-wave mixing},\ }\href
  {https://doi.org/10.1103/PhysRevApplied.19.044056} {\bibfield  {journal}
  {\bibinfo  {journal} {Phys. Rev. Appl.}\ }\textbf {\bibinfo {volume} {19}},\
  \bibinfo {pages} {044056} (\bibinfo {year} {2023})}\BibitemShut {NoStop}%
\bibitem [{\citenamefont {Kapur}\ \emph {et~al.}(2026)\citenamefont {Kapur},
  \citenamefont {Kehayias}, \citenamefont {Tolpygo}, \citenamefont {Libson},
  \citenamefont {Haldeman}, \citenamefont {Muniz}, \citenamefont {Wynn},
  \citenamefont {O'Connor}, \citenamefont {Parmar}, \citenamefont {Johnson},
  \citenamefont {Maccabe}, \citenamefont {Cummings}, \citenamefont {Mallek},
  \citenamefont {Braje},\ and\ \citenamefont {Schloss}}]{Kapur2025}%
  \BibitemOpen
  \bibfield  {author} {\bibinfo {author} {\bibfnamefont {R.~T.}\ \bibnamefont
  {Kapur}}, \bibinfo {author} {\bibfnamefont {P.}~\bibnamefont {Kehayias}},
  \bibinfo {author} {\bibfnamefont {S.~K.}\ \bibnamefont {Tolpygo}}, \bibinfo
  {author} {\bibfnamefont {A.~A.}\ \bibnamefont {Libson}}, \bibinfo {author}
  {\bibfnamefont {G.}~\bibnamefont {Haldeman}}, \bibinfo {author}
  {\bibfnamefont {C.~N.}\ \bibnamefont {Muniz}}, \bibinfo {author}
  {\bibfnamefont {A.}~\bibnamefont {Wynn}}, \bibinfo {author} {\bibfnamefont
  {N.~J.}\ \bibnamefont {O'Connor}}, \bibinfo {author} {\bibfnamefont {N.~A.}\
  \bibnamefont {Parmar}}, \bibinfo {author} {\bibfnamefont {R.}~\bibnamefont
  {Johnson}}, \bibinfo {author} {\bibfnamefont {A.~C.}\ \bibnamefont
  {Maccabe}}, \bibinfo {author} {\bibfnamefont {J.}~\bibnamefont {Cummings}},
  \bibinfo {author} {\bibfnamefont {J.~L.}\ \bibnamefont {Mallek}}, \bibinfo
  {author} {\bibfnamefont {D.~A.}\ \bibnamefont {Braje}},\ and\ \bibinfo
  {author} {\bibfnamefont {J.~M.}\ \bibnamefont {Schloss}},\ }\bibfield
  {title} {\bibinfo {title} {Flux-trapping characterization for superconducting
  electronics using a cryogenic widefield {N}-$v$ diamond microscope},\ }\href
  {https://doi.org/10.1103/3zcp-lsrd} {\bibfield  {journal} {\bibinfo
  {journal} {Phys. Rev. Appl.}\ }\textbf {\bibinfo {volume} {25}},\ \bibinfo
  {pages} {044073} (\bibinfo {year} {2026})},\ \Eprint
  {https://arxiv.org/abs/2506.01906} {arXiv:2506.01906 [cond-mat.supr-con]}
  \BibitemShut {NoStop}%
\bibitem [{\citenamefont {Shulga}\ \emph {et~al.}(2025)\citenamefont {Shulga},
  \citenamefont {Nishimura}, \citenamefont {Volkov}, \citenamefont {Hasegawa},
  \citenamefont {Hirano}, \citenamefont {Tsuji}, \citenamefont {Iwasaki},
  \citenamefont {Hatano}, \citenamefont {Sasaki},\ and\ \citenamefont
  {Kobayashi}}]{Shulga2025}%
  \BibitemOpen
  \bibfield  {author} {\bibinfo {author} {\bibfnamefont {K.}~\bibnamefont
  {Shulga}}, \bibinfo {author} {\bibfnamefont {S.}~\bibnamefont {Nishimura}},
  \bibinfo {author} {\bibfnamefont {P.~A.}\ \bibnamefont {Volkov}}, \bibinfo
  {author} {\bibfnamefont {R.}~\bibnamefont {Hasegawa}}, \bibinfo {author}
  {\bibfnamefont {M.}~\bibnamefont {Hirano}}, \bibinfo {author} {\bibfnamefont
  {T.}~\bibnamefont {Tsuji}}, \bibinfo {author} {\bibfnamefont
  {T.}~\bibnamefont {Iwasaki}}, \bibinfo {author} {\bibfnamefont
  {M.}~\bibnamefont {Hatano}}, \bibinfo {author} {\bibfnamefont
  {K.}~\bibnamefont {Sasaki}},\ and\ \bibinfo {author} {\bibfnamefont
  {K.}~\bibnamefont {Kobayashi}},\ }\href@noop {} {\bibinfo {title}
  {Observation of individual vortex penetration in a coplanar superconducting
  resonator}} (\bibinfo {year} {2025}),\ \Eprint
  {https://arxiv.org/abs/2512.00790} {arXiv:2512.00790 [cond-mat.supr-con]}
  \BibitemShut {NoStop}%
\bibitem [{\citenamefont {Armstrong}\ \emph {et~al.}(1962)\citenamefont
  {Armstrong}, \citenamefont {Bloembergen}, \citenamefont {Ducuing},\ and\
  \citenamefont {Pershan}}]{Armstrong1962}%
  \BibitemOpen
  \bibfield  {author} {\bibinfo {author} {\bibfnamefont {J.~A.}\ \bibnamefont
  {Armstrong}}, \bibinfo {author} {\bibfnamefont {N.}~\bibnamefont
  {Bloembergen}}, \bibinfo {author} {\bibfnamefont {J.}~\bibnamefont
  {Ducuing}},\ and\ \bibinfo {author} {\bibfnamefont {P.~S.}\ \bibnamefont
  {Pershan}},\ }\bibfield  {title} {\bibinfo {title} {Interactions between
  light waves in a nonlinear dielectric},\ }\href
  {https://doi.org/10.1103/physrev.127.1918} {\bibfield  {journal} {\bibinfo
  {journal} {Phys. Rev.}\ }\textbf {\bibinfo {volume} {127}},\ \bibinfo {pages}
  {1918} (\bibinfo {year} {1962})}\BibitemShut {NoStop}%
\bibitem [{\citenamefont {Zorin}(2021)}]{Zorin2021}%
  \BibitemOpen
  \bibfield  {author} {\bibinfo {author} {\bibfnamefont {A.~B.}\ \bibnamefont
  {Zorin}},\ }\bibfield  {title} {\bibinfo {title} {Quasi-phasematching in a
  poled {J}osephson traveling-wave parametric amplifier with three-wave
  mixing},\ }\href {https://doi.org/10.1063/5.0050787} {\bibfield  {journal}
  {\bibinfo  {journal} {Appl. Phys. Lett.}\ }\textbf {\bibinfo {volume}
  {118}},\ \bibinfo {pages} {222601} (\bibinfo {year} {2021})}\BibitemShut
  {NoStop}%
\bibitem [{\citenamefont {Gandorfer}\ \emph {et~al.}(2025)\citenamefont
  {Gandorfer}, \citenamefont {Renger}, \citenamefont {Yam}, \citenamefont
  {Fesquet}, \citenamefont {Marx}, \citenamefont {Gross},\ and\ \citenamefont
  {Fedorov}}]{Gandorfer2025}%
  \BibitemOpen
  \bibfield  {author} {\bibinfo {author} {\bibfnamefont {S.}~\bibnamefont
  {Gandorfer}}, \bibinfo {author} {\bibfnamefont {M.}~\bibnamefont {Renger}},
  \bibinfo {author} {\bibfnamefont {W.~K.}\ \bibnamefont {Yam}}, \bibinfo
  {author} {\bibfnamefont {F.}~\bibnamefont {Fesquet}}, \bibinfo {author}
  {\bibfnamefont {A.}~\bibnamefont {Marx}}, \bibinfo {author} {\bibfnamefont
  {R.}~\bibnamefont {Gross}},\ and\ \bibinfo {author} {\bibfnamefont {K.~G.}\
  \bibnamefont {Fedorov}},\ }\bibfield  {title} {\bibinfo {title}
  {Two-dimensional {P}lanck spectroscopy for microwave photon calibration},\
  }\href {https://doi.org/10.1103/PhysRevApplied.23.024064} {\bibfield
  {journal} {\bibinfo  {journal} {Phys. Rev. Appl.}\ }\textbf {\bibinfo
  {volume} {23}},\ \bibinfo {pages} {024064} (\bibinfo {year}
  {2025})}\BibitemShut {NoStop}%
\bibitem [{\citenamefont {Callen}\ and\ \citenamefont
  {Welton}(1951)}]{Callen1951}%
  \BibitemOpen
  \bibfield  {author} {\bibinfo {author} {\bibfnamefont {H.~B.}\ \bibnamefont
  {Callen}}\ and\ \bibinfo {author} {\bibfnamefont {T.~A.}\ \bibnamefont
  {Welton}},\ }\bibfield  {title} {\bibinfo {title} {Irreversibility and
  generalized noise},\ }\href {https://doi.org/10.1103/physrev.83.34}
  {\bibfield  {journal} {\bibinfo  {journal} {Phys. Rev.}\ }\textbf {\bibinfo
  {volume} {83}},\ \bibinfo {pages} {34} (\bibinfo {year} {1951})}\BibitemShut
  {NoStop}%
\bibitem [{\citenamefont {Caves}(1982)}]{Caves1982}%
  \BibitemOpen
  \bibfield  {author} {\bibinfo {author} {\bibfnamefont {C.~M.}\ \bibnamefont
  {Caves}},\ }\bibfield  {title} {\bibinfo {title} {Quantum limits on noise in
  linear amplifiers},\ }\href {https://doi.org/10.1103/PhysRevD.26.1817}
  {\bibfield  {journal} {\bibinfo  {journal} {Phys. Rev. D}\ }\textbf {\bibinfo
  {volume} {26}},\ \bibinfo {pages} {1817} (\bibinfo {year}
  {1982})}\BibitemShut {NoStop}%
\bibitem [{\citenamefont {Renger}\ \emph {et~al.}(2021)\citenamefont {Renger},
  \citenamefont {Pogorzalek}, \citenamefont {Chen}, \citenamefont {Nojiri},
  \citenamefont {Inomata}, \citenamefont {Nakamura}, \citenamefont {Partanen},
  \citenamefont {Marx}, \citenamefont {Gross}, \citenamefont {Deppe},\ and\
  \citenamefont {Fedorov}}]{Renger2021}%
  \BibitemOpen
  \bibfield  {author} {\bibinfo {author} {\bibfnamefont {M.}~\bibnamefont
  {Renger}}, \bibinfo {author} {\bibfnamefont {S.}~\bibnamefont {Pogorzalek}},
  \bibinfo {author} {\bibfnamefont {Q.}~\bibnamefont {Chen}}, \bibinfo {author}
  {\bibfnamefont {Y.}~\bibnamefont {Nojiri}}, \bibinfo {author} {\bibfnamefont
  {K.}~\bibnamefont {Inomata}}, \bibinfo {author} {\bibfnamefont
  {Y.}~\bibnamefont {Nakamura}}, \bibinfo {author} {\bibfnamefont
  {M.}~\bibnamefont {Partanen}}, \bibinfo {author} {\bibfnamefont
  {A.}~\bibnamefont {Marx}}, \bibinfo {author} {\bibfnamefont {R.}~\bibnamefont
  {Gross}}, \bibinfo {author} {\bibfnamefont {F.}~\bibnamefont {Deppe}},\ and\
  \bibinfo {author} {\bibfnamefont {K.~G.}\ \bibnamefont {Fedorov}},\
  }\bibfield  {title} {\bibinfo {title} {Beyond the standard quantum limit for
  parametric amplification of broadband signals},\ }\href
  {https://doi.org/10.1038/s41534-021-00495-y} {\bibfield  {journal} {\bibinfo
  {journal} {npj Quantum Inf.}\ }\textbf {\bibinfo {volume} {7}},\ \bibinfo
  {pages} {160} (\bibinfo {year} {2021})}\BibitemShut {NoStop}%
\end{thebibliography}%

\newpage
\appendix

\section{Noise characterization}
\label{app:noise}

The JTWPA sample is measured in different dilution refrigerators and setups. An initial characterization is done in a commercial Bluefors LD400 fridge. There, we use a Keysight PNA vector network analyzer to measure $S_{21}$ scattering parameters and a vector microwave generator R\&S SGS100A RF source for controlling the pump tone  coupled to the JTWPA input via a broadband directional coupler. We use this setup to measure the JTWPA insertion losses of \qty{0.5}{\deci\bel} in the \qtyrange{4}{8}{\giga\hertz} range using a bypass microwave cryogenic switch. 

\begin{figure}
    \centering
    \includegraphics[width=1.0\linewidth]{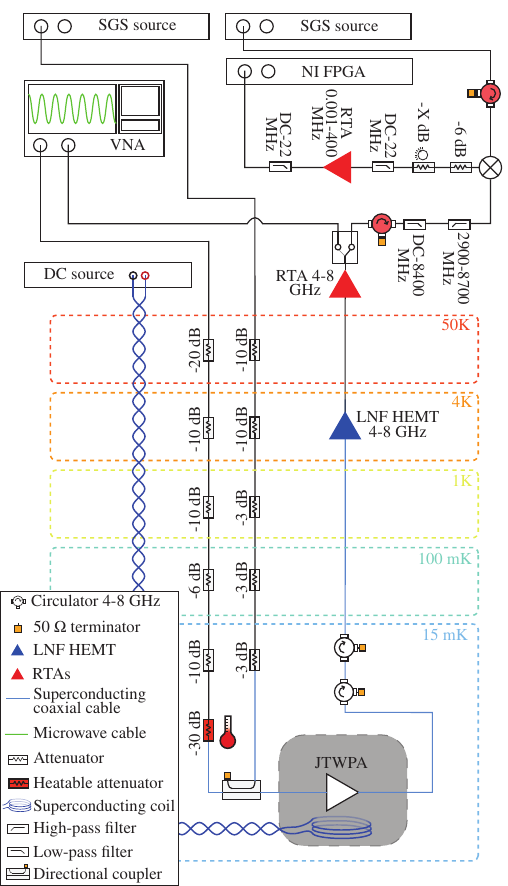}
    \caption{Schematic of the experimental setup for PNCF measurements for noise characterization. Later we modified the setup for high-frequency measurements by replacing the HEMT with an LNF \qtyrange{6}{20}{\giga\hertz} amplifier and the room-temperature \qtyrange{4}{8}{\giga\hertz} amplifier  with a BZT high-frequency amplifier having frequency range \qtyrange{6}{18}{\giga\hertz}.}
    \label{fig:app_measurement_setup}
\end{figure}

For more detailed measurements, we move the JTWPA to an Oxford Instruments Triton fridge with a \qtyrange{4}{8}{\giga\hertz} measurement setup. It also includes a heatable attenuator embedded into the JTWPA input line with a corresponding PID temperature control, a room temperature down-conversion chain and a National Instruments field-programmable gate array (FPGA) with a corresponding digitizer, as shown in Fig.\,\ref{fig:app_measurement_setup}. These additional elements allow to perform noise characterization measurements of our parametric amplifier. For the broadband transmission measurements discussed in the main text, this setup was modified for high-frequency operation as described in the caption of Fig.\,\ref{fig:app_measurement_setup}.

We use the setup depicted in Fig.\,\ref{fig:app_measurement_setup} to measure added noise during the parametric amplification in our JTWPA. Here, we rely on a heatable attenuator as a calibrated photon source obeying the Planck distribution. We sweep the attenuator temperature using the PID control implemented with a Lakeshore 372 AC room temperature resistance bridge. We use the FPGA preceded by the down-conversion stage to digitize, filter, and process microwave signals. The signal processing step includes calculation of first- and second-order signal statistical moments and allows for estimation of respective photon numbers, which, in combination with the proper calibration of a photon number conversion factor (PNCF) of our amplification chain, allows for \emph{in situ} measurement of added noise in the JTWPA.  

In particular, we rely on the 2D PNCF method~\cite{Gandorfer2025} to estimate losses of the components, mounted on the mixing chamber stage. Since we use a broadband photon source to characterize a broadband amplifier, we use a two-mode model, taking into account added noise from the idler mode\,\cite{Ranadive2022, Gaydamachenko2025, Malnou2021}.

We use Planck's law in the Callen-Welton form to calculate the photon number at frequency $f$ from a source with temperature $T$~\cite{Callen1951}
\begin{equation}
    N(f, T) = \frac{1}{2}\coth\left({\frac{hf}{2k_\mathrm{B}T}}\right)\,,
    \label{eq:Planck}
\end{equation}
where $h$ is the Planck constant and $k_\mathrm{B}$ is the Boltzmann constant. We use this equation to calculate the power emitted by the heatable attenuator in front of the JTWPA and also the amount of noise photons coupled from the thermal bath.

To take into account losses, we treat the lossy element as an asymmetric beam splitter with transmittivity $\eta$
\begin{equation}
   N^{\mathrm{BS}}(N_\mathrm{in}, \eta, f, T_\mathrm{bath}) = \eta N_\mathrm{in} + (1-\eta) N_\mathrm{bath}(f, T_\mathrm{bath}).
   \label{eq:beam-splitter-model}
\end{equation}
Here, the input signal $N_\mathrm{in}$ is attenuated by a factor $\eta$, and the mean photon number of the environment with temperature $T_\mathrm{bath}$ couples to the microwave channel.

Next, we treat the JTWPA as a mixer with two input ports and one output port, to take into account the broadband nature of the amplifier and the noise source. As mentioned in Refs.\,\cite{Ranadive2022,Malnou2021}, using a single mode model for this case will lead to underestimation of actual added noise up to two times.

In this model, the output JTWPA power at frequency $f_\mathrm{s}$ is combined from amplified signal and idler modes $N_\mathrm{in}^\mathrm{s}$ and $N_\mathrm{in}^i$ at frequencies $f_\mathrm{s}$ and $f_\mathrm{i}$, emitted by the broadband noise source, with corresponding numbers of added noise photons $n_\mathrm{s}$ and $n_\mathrm{i}$

\begin{multline}
    N^\mathrm{TWPA}_\mathrm{out} = G_\mathrm{s}\biggl(N^\mathrm{BS}(N_\mathrm{in}^\mathrm{s}, \eta_\mathrm{s}, f_\mathrm{s}, T_\mathrm{MXC}) + n_\mathrm{s}\biggr) +  \\
    (G_\mathrm{s}-1)\biggl(N^\mathrm{BS}(N_\mathrm{in}^\mathrm{i}, \eta_\mathrm{i}, f_\mathrm{i}, T_\mathrm{MXC}) + n_\mathrm{i}\biggr)\,.
\end{multline}

Note, that here we took into account losses between the heatable attenuator and the JTWPA input $\eta_\mathrm{s}$ and $\eta_\mathrm{i}$, corresponding to losses at the signal and idler frequencies and calculated via the beam-splitter model (Eq.\,\ref{eq:beam-splitter-model}) assuming the temperature of the lossy components is equal to the mixing chamber stage temperature $T_\mathrm{MXC}$.

Next, we apply the beam-splitter model to take into account losses after the TWPA $\eta_\mathrm{after}$ to recalculate the number of photons into power that we acquire with the room temperature electronics
\begin{multline}
    P(f_\mathrm{s}, T_\mathrm{att}, T_\mathrm{MXC}) = \frac{\kappa}{Z_0}\biggl(N^\mathrm{BS}(N^\mathrm{TWPA}_\mathrm{out}, \eta_\mathrm{after},T_\mathrm{MXC}, f_\mathrm{s}) \\ 
    + n_\mathrm{HEMT}\biggr)\,.
    \label{eq:PNCF-power}
\end{multline}
Here, the photon number conversion factor $\kappa = G_\mathrm{4K-RT} Z_0Bhf$ takes into account the gain of the amplifying chain after the JTWPA $G_\mathrm{4K-RT}$, detection bandwidth $B$, and characteristic impedance $Z_0=\qty{50}{\ohm}$. 

\begin{figure}
    \centering
    \includegraphics[width=1.0\linewidth]{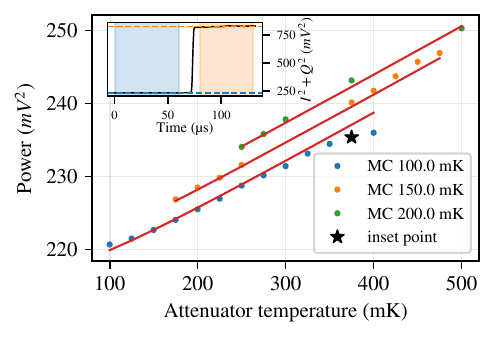}
    \caption{2D PNCF fitting procedure for pump-off case for the signal at \qty{4}{\giga\hertz}. We change the temperature of the heatable attenuator for different temperatures of the mixing chamber flange (\qtylist{100;150;200}{\milli\kelvin}) and measure the power at a given frequency. We show fit lines via Eq.~\ref{eq:PNCF-power} as solid lines. \textit{In the inset:} raw data for the point, marked by a star on the main plot. We measure incoming power for \qty{130}{\micro\second} at each point (MXC temperature, attenuator temperature, signal frequency, pump frequency), applying the pump pulse for the second half of the measuring window. In post-processing we average the signal to get pump-on and pump-off point to obtain plots, similar to the main plot.}
    \label{fig:app_PNCF}
\end{figure}

\begin{figure}
    \centering
    \includegraphics[width=1.0\linewidth]{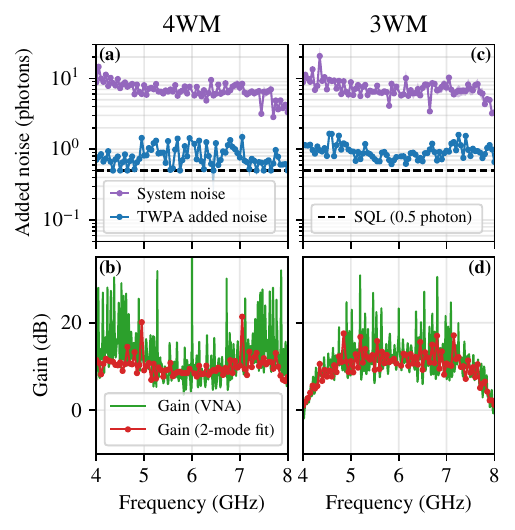}
    \caption{Noise performance of the device operated at the optimal points for 4WM and 3WM. (a) Added system noise (purple) measured with JTWPA pump off and added JTWPA noise (blue) measured with pump on for the 4WM operation mode. (b) 4WM gain profile in the optimal point measured with the VNA (green) and extracted from PNCF fits (red). (c) Added system noise (purple) measured with JTWPA pump off and added JTWPA noise (blue) measured with pump on for the 3WM operation mode. (d) 3WM gain profile in the optimal point measured with the VNA (green) and extracted from PNCF fits (red).}
    \label{fig:app_noise_photons}
\end{figure}

The complete noise characterization protocol is as follows: first, after setting current through the magnetic coil and pump parameters to the optimal values, we fix the mixing chamber temperature.  In the next step, we set the heatable attenuator to the highest temperature of the sweep. After waiting for 10 minutes for the components to thermalize, we perform PNCF measurements for each carrier frequency, which we set by applying a corresponding frequency $f_\mathrm{s} + f_\mathrm{IF}$, where $f_\mathrm{IF}=\qty{12.5}{\mega\hertz}$ is the intermediate frequency, to the mixer at room temperature (see Fig.\,\ref{fig:app_measurement_setup}). We measure moments with the FPGA up to the third order for \qty{130}{\micro\second}. In the first \qty{65}{\micro\second} we measure the signal coming from unpumped JTWPA, followed by \qty{65}{\micro\second} of acquiring the signal with the pump on. After this, we set a lower temperature of the heatable attenuator and repeat the measurements after waiting for thermalization. We conduct these measurements for different temperatures of the MXC flange, which allows to estimate losses of components mounted at this flange\,\cite{Gandorfer2025}.

Then, we perform analysis of the dataset. By averaging $\left<I^2\right>+\left<Q^2\right>$ for the first half of the pulse (see inset in Fig.\,\ref{fig:app_PNCF}) and fitting datasets acquired at different temperatures of the mixing chamber stage with the 2D PNCF approach we extract losses at the mixing chamber stage and system noise photons, mostly defined by HEMT performance.

The next step is to analyze the dataset of the pumped JTWPA. For this we use only data for the lowest temperature of the mixing chamber as well as losses and added noise from the 2D PNCF of unpumped data. From the analysis of the pumped JTWPA dataset using the two-mode model, we extract added noise for each carrier frequency. 

We characterize the JTWPA's noise performance for optimal parameters, resulting in the largest gain in 4WM and 3WM regimes, as shown for the two starred points in Fig.\,\ref{fig:3WM_4WM_avg_gain_maps}. We present added amplifier noise $n_\mathrm{s}+n_\mathrm{i}$ as a function of frequency in Fig.\,\ref{fig:app_noise_photons}(a,c). The plotted uncertainty is estimated from uncertainties of consecutive fits of PNCF values, losses at mixing chamber, added noise photons from HEMT and losses ratio ($\eta_\mathrm{after}/\eta_\mathrm{s,(i)}$). We estimate added noise to be $\num{0.32}\pm\num{0.27}$ and  $\num{0.45}\pm\num{0.23}$  photons on average above the SQL for 4WM and 3WM operation regimes, respectively. Here, we define the SQL as $\num{0.5}$ photons according to Ref.\,\cite{Caves1982}. In principle, for broadband input signals, the SQL can be reduced to 0\,\cite{Renger2021}, however in this case we consider the idler not to carry a signal in the intended narrowband application, such as increasing the SNR of dispersive readout of a superconducting processor.

We additionally compare the gain measured with the VNA to that extracted from the fitting routine, given by $G = \kappa_{\mathrm{on}}/\kappa_{\mathrm{off}}$, as shown in Fig.\,\ref{fig:app_noise_photons}(b) and \ref{fig:app_noise_photons}(d). Both methods show good agreement in terms of gain magnitude and spectral shape. 

\section{Three-mode model}
\label{app:three-mode-model}

We model the device as a discrete transmission line formed by
$N_\mathrm{cells}=\num{912}$ identical SNAIL cells. Each cell consists of one small Josephson junction with critical current $\alpha I_\mathrm{c}$ in parallel with an array of three large junctions, each having critical current $I_\mathrm{c}$. For a local magnetic flux $\Phi_n$ through cell $n$, the normalized SNAIL potential and current--phase relation are
\begin{align}
    u_n(\delta)
    &=
    -\alpha\cos\delta
    -3\cos\left(\frac{\varphi_n-\delta}{3}\right)\,,
    \label{eq:snail_potential}
    \\
    I_n(\delta)
    &=
    \alpha I_\mathrm{c}\sin\delta
    -I_\mathrm{c}\sin\left(\frac{\varphi_n-\delta}{3}\right)\,,
\label{eq:snail_current}
\end{align}
where $u_n=U_n/E_\mathrm{J}$ is the potential normalized by $E_\mathrm{J}=\varphi_0 I_\mathrm{c}$, $\varphi_0=\Phi_0/(2\pi)$ is the reduced flux quantum, $\Phi_0$ is the superconducting flux quantum, $\varphi_n=2\pi\Phi_n/\Phi_0$ is the reduced external flux, and
$\delta$ is the phase drop across the small junction. At every local flux, the static phase $\delta_{0,n}$ is obtained by numerical minimization of Eq.~(\ref{eq:snail_potential}). The current is then expanded about this minimum as
\begin{equation}
    I_n(\delta_{0,n}+x)
    =
    i_{1,n}x
    +\frac{i_{2,n}}{2}x^2
    +\frac{i_{3,n}}{6}x^3
    +\mathcal{O}(x^4)\,,
\label{eq:snail_expansion}
\end{equation}
where $x=\delta-\delta_{0,n}$ and
\begin{align}
    i_{1,n}
    &=
    \alpha I_\mathrm{c}\cos\delta_{0,n}
    +\frac{I_\mathrm{c}}{3}
     \cos\left(\frac{\varphi_n-\delta_{0,n}}{3}\right)\,,
    \\
    i_{2,n}
    &=
    -\alpha I_\mathrm{c}\sin\delta_{0,n}
    +\frac{I_\mathrm{c}}{9}
     \sin\left(\frac{\varphi_n-\delta_{0,n}}{3}\right)\,,
    \\
    i_{3,n}
    &=
    -\alpha I_\mathrm{c}\cos\delta_{0,n}
    -\frac{I_\mathrm{c}}{27}
     \cos\left(\frac{\varphi_n-\delta_{0,n}}{3}\right)\,.
\end{align}
Thus, $i_{1,n}$ determines the linear Josephson inductance $L_n=\varphi_0/i_{1,n}$, whereas $i_{2,n}$ and $i_{3,n}$ determine the local three-wave- and four-wave-mixing nonlinearities, respectively. The simulations use $\alpha=\num{0.12}$, $I_\mathrm{c}=\qty{2.1}{\micro\ampere}$, and a large-junction capacitance $C_\mathrm{J}=\qty{100}{\femto\farad}$. The effective inter-node capacitance of one SNAIL is approximated by $C_{\mathrm{S}}=\alpha C_\mathrm{J}+C_\mathrm{J}/3=\qty{45.3}{\femto\farad}$.

The magnetic field is spatially nonuniform and is sampled at the positions $(x_n,y_n)$ of the SNAILs along the meander. The local flux used is the same as in the main text, defined through Eq.~\ref{eq:field_distribution} and shown in Fig.~\ref{fig:magnetic_field_distribution}(b). 

The resulting profile is normalized to the maximum value sampled by the meander, and the proportionality between the coil current and the peak flux $\Phi_\mathrm{max}$ of Eq.\,\ref{eq:field_distribution} is treated as an independently adjustable coil calibration.

Each SNAIL is represented by the parallel admittance of $L_n$, $C_{\mathrm{S}}$, and an effective conductance, while every node is coupled to ground through $C_\mathrm{g}=\qty{100}{\femto\farad}$. Both ends are terminated by $Z_0=\qty{50}{\ohm}$. The pump phase drop
$\Delta_{\mathrm{p},n}$ is obtained by solving the full spatially dependent nodal problem. At the pump angular frequency $\omega_\mathrm{p}$, the branch admittance is
\begin{equation}
    Y_{\mathrm{p},n}=G_n+\mathrm{i}\omega_\mathrm{p} C_{\mathrm{S}} + \frac{D_{0,n}}{\mathrm{i}\omega_\mathrm{p}\varphi_0}\,,\qquad D_{0,n}=i_{1,n}+\frac{i_{3,n}}{4}|\Delta_{\mathrm{p},n}|^2\,,
    \label{eq:pump_admittance}
\end{equation}
where $D_{0,n}$ includes the pump-induced Kerr correction and $G_n$ is the total effective conductance. After solving for the node voltages $V_{\mathrm{p},n}$, the pump phase is updated according to
\begin{equation}
    \Delta_{\mathrm{p},n}=\frac{V_{\mathrm{p},n}-V_{\mathrm{p},n+1}}{\mathrm{i}\omega_\mathrm{p}\varphi_0}\,.
\end{equation}
The nonlinear pump profile is found by four fixed-point iterations. The input pump power is \qty{-74}{\deci\belmilliwatt}, referred to a \qty{50}{\ohm} source. Pump depletion by the weak signal and idler is neglected.

Linear loss is introduced through
\begin{equation}
    G_{0,n}=\frac{1}{Q_i\sqrt{L_n/C_{\mathrm{S}}}}\,,
\end{equation}
where $Q_i=200$ is an effective cell quality factor. Leakage into harmonics and sidebands excluded from the reduced harmonic-balance basis is represented phenomenologically by
\begin{equation}
    G_{\mathrm{leak},n}=\frac{\kappa_{\mathrm{leak}}\,w_n\,|\Delta_{\mathrm{p},n}|^2}{\sqrt{L_n/C_{\mathrm{S}}}}\,,
\label{eq:leakage}
\end{equation}
with $\kappa_{\mathrm{leak}}=\num{0.02}$. Here, $w_n$ enhances the loss in strongly nonlinear cells,
\begin{equation}
    w_n = 1+\kappa_{\mathrm{nl}}\left[\frac{1}{2}\left(\frac{|i_{2,n}|}{i_{2,\mathrm{ref}}}+\frac{|i_{3,n}|}{i_{3,\mathrm{ref}}}\right)\right]^2\,,
\label{eq:nonlinear_loss_weight}
\end{equation}
where $\kappa_{\mathrm{nl}}=0.9$, while $i_{2,\mathrm{ref}}$ and $i_{3,\mathrm{ref}}$ are the maximum magnitudes of the corresponding coefficients over the simulated flux interval. Equations~(\ref{eq:leakage}) and (\ref{eq:nonlinear_loss_weight}) constitute a phenomenological description of unobserved multi-mode conversion and should not be interpreted as a microscopic dissipation model.

The weak signal is treated within a reduced harmonic-balance approximation. Three-wave mixing and four-wave mixing processes are calculated separately by retaining the signal and one conjugate idler component. Their physical idler frequencies are
\begin{equation}
    \omega_\mathrm{i}^{(3)}=\omega_\mathrm{p}-\omega_\mathrm{s}\,,\qquad\omega_\mathrm{i}^{(4)}=2\omega_\mathrm{p}-\omega_\mathrm{s}\,,
\end{equation}
where $\omega_\mathrm{s}$, $\omega_\mathrm{p}$, and $\omega_\mathrm{i}$ are the signal, pump, and idler angular frequencies, respectively. For either process $q=3,4$, the local two-frequency branch admittance has the form
\begin{equation}
    \mathbf{Y}^{(q)}_n\!\!=\!\!\begin{bmatrix}
        G_n+\mathrm{i}\omega_\mathrm{s}C_{\mathrm{S}}
        +\dfrac{D_{0,n}}{\mathrm{i}\omega_\mathrm{s}\varphi_0}
        &
        \dfrac{D_{q,n}}{\mathrm{i}\widetilde{\omega}^{(q)}_\mathrm{i}\varphi_0}
        \\[8pt]
        \dfrac{D_{q,n}^{*}}{\mathrm{i}\omega_\mathrm{s}\varphi_0}
        &
        G_n+\mathrm{i}\widetilde{\omega}^{(q)}_\mathrm{i}C_{\mathrm{S}}
        +\dfrac{D_{0,n}}
        {\mathrm{i}\widetilde{\omega}^{(q)}_\mathrm{i}\varphi_0}
    \end{bmatrix}\,,
\label{eq:hb_matrix}
\end{equation}
where $\widetilde{\omega}^{(3)}_\mathrm{i}=\omega_\mathrm{s}-\omega_\mathrm{p}$ and
$\widetilde{\omega}^{(4)}_\mathrm{i}=\omega_\mathrm{s}-2\omega_\mathrm{p}$ are signed negative-frequency representations of the conjugate idler, the asterisk denotes complex conjugation, and the mixing coefficients are
\begin{equation}
    D_{3,n}
    =
    \frac{1}{2}\,i_{2,n}\Delta_{\mathrm{p},n}\,,
    \qquad
    D_{4,n}
    =
    \frac{1}{8}i_{3,n}\Delta_{\mathrm{p},n}^{\,2}\,.
\label{eq:mixing_coefficients}
\end{equation}
The factors $1/2$ and $1/8$ follow directly from the current expansions.

The Kirchhoff equations for the $N_\mathrm{cells}+1$ nodes form a block-tridiagonal linear system, which is solved using a block Thomas algorithm. For each coil current and signal frequency, the pumped output voltage is normalized by the output of the same line calculated without the
pump,
\begin{equation}
    \mathcal{G}_q=20\log_{10}\left|\frac{V_{\mathrm{out},q}}{V_{\mathrm{out},0}}\right|\,,
\end{equation}
where $V_{\mathrm{out},q}$ is the output signal voltage obtained for process $q$ and $V_{\mathrm{out},0}$ is the unpumped output voltage. For visualization, the combined map is defined pointwise as
\begin{equation}
    \mathcal{G}_{\mathrm{comb}}
    =
    \max\left(
    \mathcal{G}_{3},
    \mathcal{G}_{4}
    \right)\,.
\end{equation}
This final operation is not a coherent combination of 3WM and 4WM. The model is therefore intended to reproduce the locations and qualitative shapes of separated amplification regions. It neglects pump depletion, signal saturation, coherent interference between 3WM
and 4WM, cascaded conversion, and explicit higher pump harmonics.

\section{Multi-mode model}
\label{app:multi-mode-model}

For the multi-mode model, we consider a transmission line formed by $N_\mathrm{cells}=\num{912}$ unit cells. Each unit cell consists of two electrical nodes with symmetrically distributed capacitances to ground and a SNAIL nonlinearity connecting the two nodes. The latter is modeled by a third-order Taylor expansion about the minimum-energy phase of the SNAIL. The complete transmission line, therefore, comprises $N_\mathrm{cells}+1$ electrical nodes. We denote the phase difference across the $j$-th SNAIL branch, connecting nodes $j$ and $j+1$, by $\varphi_{\mathrm{J},j}=\varphi_{j+1}-\varphi_j$. The current $i_j(\varphi_{\mathrm{J},j})$ through the $j$-th SNAIL branch is then given by
\begin{equation}
    i_j(\varphi_{\mathrm{J},j})=\frac{2\pi}{\Phi_0}\left[c_{2,j}\varphi_{\mathrm{J},j}+c_{3,j}\varphi_{\mathrm{J},j}^2+c_{4,j}\varphi_{\mathrm{J},j}^3\right]\, ,
\end{equation}
where $c_{2,j}$, $c_{3,j}$, and $c_{4,j}$ denote the inverse-inductance coefficient, the 3WM coefficient, and the 4WM coefficient of the $j$-th unit cell, respectively.

Ultimately, we seek the phases $\varphi_j$ at all electrical nodes $j$. To capture the relevant multi-mode mixing products within a Harmonic Balance ansatz, we represent the node phases by a superposition of harmonic contributions in terms of a Fourier series. The only externally imposed frequency references are the pump frequency $\omega_\mathrm{p}$ and the signal frequency $\omega_\mathrm{s}$. All mixing frequencies $\omega_\mathrm{mix}$ generated through 3WM and 4WM interactions can therefore be written as integer combinations of the pump and signal frequencies,
\begin{equation}
    \omega_\mathrm{mix}=n_\mathrm{p}\omega_\mathrm{p}+n_\mathrm{s}\omega_\mathrm{s}\, ,
\end{equation}
with $n_\mathrm{p},n_\mathrm{s}\in\mathbb{Z}$. Hence, the phase at node $j$ can be represented by the Fourier series
\begin{equation}
    \varphi_j=\sum\limits_{n_\mathrm{p}=-\infty}^{\infty}\sum\limits_{n_\mathrm{s}=-\infty}^{\infty}c_{n_\mathrm{p},n_\mathrm{s},j}\e^{\imu\left(n_\mathrm{p}\omega_\mathrm{p}+n_\mathrm{s}\omega_\mathrm{s}\right)t}\, .
\end{equation}
The complex Fourier coefficients $c_{n_\mathrm{p},n_\mathrm{s},j}$ determine the spatial distribution of the phase along the JTWPA for each retained harmonic contribution. Since $\varphi_j$ is real valued, the Fourier coefficients additionally satisfy the conjugate-symmetry relation $c_{-n_\mathrm{p},-n_\mathrm{s},j}=c_{n_\mathrm{p},n_\mathrm{s},j}^{\ast}$. The contribution to a harmonic identified by the integer pair $\left(k,l\right)$ arising from the quadratic term associated with 3WM is given by the convolution
\begin{equation}
    \left[\varphi_j^2\right]_{\left(k,l\right)}=\sum\limits_{n_\mathrm{p}=-\infty}^{\infty}\sum\limits_{n_\mathrm{s}=-\infty}^{\infty}c_{n_\mathrm{p},n_\mathrm{s},j}c_{k-n_\mathrm{p},l-n_\mathrm{s},j}\, .
\end{equation}
The corresponding cubic contribution associated with 4WM is obtained from the repeated convolution
\begin{equation}
    \left[\varphi_j^3\right]_{\left(k,l\right)}=\sum\limits_{n_\mathrm{p}=-\infty}^{\infty}\sum\limits_{n_\mathrm{s}=-\infty}^{\infty}\left[\varphi_j^2\right]_{\left(n_\mathrm{p},n_\mathrm{s}\right)}c_{k-n_\mathrm{p},l-n_\mathrm{s},j}\, .
\end{equation}
For the numerical implementation, we restrict the harmonic expansion to a finite frequency basis by introducing a cut-off frequency $\omega_\mathrm{max}$, which may, for example, be chosen in relation to the minimum plasma frequency of the SNAIL unit cells. In general, the frequency cut-off can be chosen freely provided that higher-frequency contributions outside the retained basis are sufficiently suppressed. In addition, we limit the number of mixing products by imposing finite bounds on the integer indices, $\left|n_\mathrm{p}\right|\leq N_\mathrm{p}$ and $\left|n_\mathrm{s}\right|\leq N_\mathrm{s}$. Only combinations satisfying the prescribed frequency and integer cut-offs are retained in the harmonic balance basis.

Boundary conditions at the left and right ends of the JTWPA are imposed at the first and last electrical nodes through ports connected to an environment with a predefined reference impedance, typically chosen as \qty{50}{\ohm}. The pump and signal tones are injected through the left port. By enforcing Kirchhoff's current law at every electrical node together with the corresponding port boundary conditions, we obtain a coupled nonlinear system of equations for the Fourier coefficients of all retained modes along the spatial unit-cell index $j$. The nonlinear mixing terms densely couple the retained frequency components within each unit cell, whereas the spatial coupling remains sparse because each electrical node is coupled only to its neighboring unit cells.

The resulting multidimensional nonlinear system of equations is solved using the Newton-Raphson algorithm. A critical aspect of the numerical solution is the choice of a suitable initial guess. We first determine a solution in the absence of pump and signal excitation. Starting from this operating point, the pump amplitude is gradually increased while the signal amplitude is kept at zero. Each converged pump-only solution is used as the initial guess for the subsequent pump amplitude until the desired pump power is reached. The resulting pump-only solution is then used as the initial guess for the signal simulations, in which a weak input signal is applied and its frequency is swept over the range \qtyrange{4}{10}{\giga\hertz}. Since the signal constitutes only a weak perturbation of the pump-driven state, the pump-only solution provides a suitable initial guess for these simulations. The gain of the JTWPA is subsequently obtained from the Fourier coefficients of the signal-frequency component at the input and output ports.

The simulations are repeated for pump frequencies of \qtylist{10;11;12}{\giga\hertz}. We choose a ground capacitance of $C_\mathrm{g}=\qty{80}{\femto\farad}$ and a substrate loss tangent of $\tan\delta=\num[group-digits=integer]{0.0025}$. The critical current $I_\mathrm{c}$ and the effective SNAIL capacitance $C_\mathrm{J}$ are chosen such that the minimum SNAIL plasma frequency is approximately \qty{18}{\giga\hertz}, while still allowing sufficiently large pump amplitudes. For the considered pump frequencies, the second pump harmonic at $2\omega_\mathrm{p}$ lies in the range \qtyrange{20}{24}{\giga\hertz}, i.e., above the minimum SNAIL plasma frequency. This choice, therefore, disfavors the propagation and buildup of the second pump harmonic and reduces its excitation via second-harmonic generation, thereby promoting efficient 3WM parametric amplification. The SNAILs in each unit cell feature $N=3$ large junctions and one small junction with an asymmetry ratio of $\alpha=\num{0.12}$. The input pump power is set to \qty{-74}{\deci\belmilliwatt}. The simulation code is implemented in a modular Python framework, enabling flexible definitions of the harmonic basis, unit-cell parameters, boundary conditions, and nonlinear solution procedure.

\end{document}